\documentclass[twocolumn,pre,aps,floatfix]{revtex4-1}

\usepackage{amsmath}
\usepackage{graphicx}
\usepackage{epstopdf}
\usepackage[usenames]{color}
\usepackage{xr}
\usepackage{placeins}
\usepackage[breaklinks=true,
            pdfborder={0 0 1},
            colorlinks=true,
            linkcolor=black,
            citecolor=blue,
            urlcolor=blue]{hyperref}
\usepackage[normalem]{ulem}
\usepackage{rotating}
\usepackage{verbatim}
\usepackage{bibentry}
\usepackage{xspace}
\usepackage[dvipsnames]{xcolor}
\usepackage{float}
\usepackage{amsmath,amssymb}

\def\kt{k_\text{B}T}

\def\Vwca{V_\text{WCA}}

\begin{document}

\title{Response of active Brownian particles to boundary driving}

\author{Caleb G. Wagner}
\email{email:kswag@brandeis.edu}
\author{Michael F. Hagan}
\email{email:hagan@brandeis.edu}
\author{Aparna Baskaran}
\email{email:aparna@brandeis.edu}

\affiliation{Martin Fisher School of Physics, Brandeis University,
  Waltham, MA, USA.}

\begin{abstract}
We computationally study the behavior of underdamped active Brownian particles in a sheared channel geometry. Due to their underdamped dynamics, the particles carry momentum a characteristic distance away from the boundary before it is dissipated into the substrate. We correlate this distance with the persistence of particle trajectories, determined jointly by their friction and self-propulsion. Within this characteristic length, we observe new and counterintuitive phenomena stemming from the interplay of activity, interparticle interactions, and the boundary driving. Depending on values of friction and self-propulsion, interparticle interactions can either aid or hinder momentum transport. More dramatically, in certain cases we observe a flow reversal near the wall, which we correlate with an induced polarization of the particle self-propulsion directions. We rationalize these results in terms of a simple kinetic picture of particle trajectories.
\end{abstract}
%
%We study numerically the behavior of underdamped active Brownian particles (ABPs) in a sheared channel geometry. Due to their underdamped dynamics, the ABPs carry momentum a characteristic distance into the bulk before it is dissipated into the substrate. We identify this characteristic length in terms of the \emph{persistence} of the particle trajectories, as determined by the friction and self-propulsion. Thus, the statistics of momentum transport is confined to a boundary layer on the scale of particle persistence. Within this boundary layer, we find new and counterintuitive phenomena. Depending on values of friction and self-propulsion, interparticle interactions can either aid or hinder momentum transport. Even more dramatically, in certain cases we observed a \emph{flow reversal} near the wall, which we correlated with an induced polarization of the direction of self-propulsion. Our results lend insight into the generic response of self-propelled particles to external driving.
%Finally, our results lend insight into the generic response of self-propelled particles to external driving.

\maketitle

\section{Introduction}

Systems which are driven far from equilibrium exhibit emergent phenomena that are strikingly different from the thermodynamically allowed behaviors of equilibrium systems. Recently, intense research has focused on a class of such systems known as active matter, in which
driving enters the system at the level of its microscopic constituents \cite{Ramaswamy2010,Marchetti2013,Bechinger2016,Zottl2016,Yoshinaga2017}. Active matter occurs on many scales,
from the microscopic and colloidal to the macroscopic. Specific examples include the cell cytoskeleton \cite{Schaller2011}, bacterial suspensions \cite{Dombrowski2004,Kaiser2014}, synthetic self-propelled colloids \cite{Palacci2010,Narayan2006a,Narayan2007,Bartolo2013,Chate2012,Dauchot2017},  schooling fish \cite{Becco2006,Cambui2012}, and flocking birds \cite{Attanasi2014}.

Progress toward a fundamental understanding of active matter requires minimal models that are sufficiently tractable to describe theoretically, but exhibit the key phenomenology of more complicated, real-world systems. Toward this end, a common paradigm is to consider particles
which self-propel as a result of an internal driving force acting along some body axis. For example, the
active Brownian particle (ABP) model describes spheres or discs that self-propel at constant
velocity and whose direction of propulsion evolves diffusively \cite{Marchetti2016}.
Despite their simplicity, such self-propelled particle models exhibit striking emergent phenomena, including athermal phase separation \cite{Fily2012,Redner2013,Redner2013b,Stenhammar2013,Buttinoni2013,
Mognetti2013,Stenhammar2014b,Wysocki2014,Cates2015,Stenhammar2015,Ni2013a},
spontaneous flows \cite{Tailleur2009,Wan2008,Angelani2009,Ghosh2013,Li2017,Reichhardt2017}, and long-range density variations \cite{Harder2014,Ni2015,Angelani2011,Solon2018,Rodenburg2018}. However, researchers have only recently begun to study these models in the presence of \emph{external} driving.  Previous work has examined the response of self-propelled particles to perturbing external fields \cite{Caprini2018} and time-periodic compression/expansion \cite{Wang2018}. Efforts have also been made to construct a formal theoretical framework of response and transport in active materials, using an Irving-Kirkwood-type approach \cite{Klymko2017,Epstein2019}, a multiple-time-scale analysis \cite{Steffenoni2017}, or large deviation theory \cite{GrandPre2018}.

%How does time-dependent response depend on its intrinsic parameters such as density and self-propulsion?
%Are there analogies with linear response theory of passive materials? What truly nonequilibrium phenomena
 %might arise as a result of the unique interplay between activity and external driving?

It has been established that boundaries have dramatic and long-ranged effects in active systems, which make active systems non-extensive (i.e. their behaviors are not independent of system size)\cite{Ni2015,Yaouen2015,Solon2015b,Wagner2017,Solon2018,Rodenburg2018,Yan2018}. However, the consequences of boundary \textit{driving} have yet to be addressed in the literature of active particles. In this article, we begin to address this question by performing computer simulations of an underdamped ABP system subject to shearing forces applied at the boundary. We characterize the response of the system in terms of the flow velocity profile -- defined as the average particle velocity at a given position -- and analyze the results in the context of a simple kinetic picture of particle trajectories.

In general, the flow velocity profile decays exponentially with distance from the boundary. We denote the length of this decay as the \emph{penetration depth}, which generically depends on the friction and self-propulsion forces. Interestingly, we find that interparticle interactions can either aid or hinder momentum transport depending on the system parameters. This stands in contrast with systems of passive spheres, where interactions generically enhance momentum transport.

In order to shed light on possible boundary conditions applicable to continuum theories of rheology of active fluids, we consider also the properties of the system \emph{at} the wall, i.e. on the order of a particle diameter from the wall. In further contrast to equilibrium systems, we discover a flow reversal phenomenon within this region, where the flow velocity points opposite to the boundary driving. Finally, we find that the stress at the wall is a nontrivial function of the density of the system.

We rationalize these findings in terms of a simple kinetic picture of how ABPs move and interact in the presence of shear stress. We conclude that the response of ABPs to boundary driving is dominated by a boundary layer on the scale of the persistence of particle trajectories. Finally, we discuss the implications of our results for developing a more
systematic theoretical description of response and transport in systems of self-propelled particles.

\section{Model and simulations}
\label{sec:model}

\emph{Equations of motion.} We work within the active Brownian particle (ABP) model, which is an
idealized model system that captures important features of several  experimental active matter systems, such as vibration-fluidized granular matter and chemically-propelled colloidal particles \cite{Palacci2010,Buttinoni2013,Dauchot2017,Walsh2017}. In general,
ABPs are self-propelled spheres with diffusive reorientation statistics. In
our case we specialize to two-dimensional systems in which the translational center-of-mass dynamics is underdamped with corresponding friction coefficient $\xi $. Physically, this can be conceptualized as particle motion on a two-dimensional dissipative substrate. On the other hand, we keep the angular dynamics overdamped, since angular inertia is expected to play only a secondary role in the transport of linear momentum. The equations of motion are then
\begin{eqnarray}
\frac{d\mathbf{r}}{dt} &=&\mathbf{v} \label{EOM-dimensional1-eq} \\
\frac{d\mathbf{v}}{dt} &=&\frac{F_p}{m}\widehat{\mathbf{u}}-\frac{1}{m} \nabla
\Vwca-\xi \mathbf{v}+\sqrt{2D}\xi \mathbf{\eta }^{T}(t)  \label{EOM-dimensional2-eq} \\
\frac{d\theta }{dt} &=&\sqrt{2D_{r}}\eta ^{R}(t). \label{EOM-dimensional3-eq}
\end{eqnarray}
Here,  $\mathbf{\eta }^{T}(t)$
and $\eta ^{R}(t)$ are delta-correlated thermal noises, i.e. satisfying $\langle \eta
(t)\eta (t^{\prime })\rangle =\delta (t-t^{\prime })$ with corresponding
diffusion coefficients $D$ and $D_{r} = 3D/\sigma^2$. The self-propulsion enters as the constant magnitude force $F_{p}$ in the
direction of a particle's orientation $\widehat{\mathbf{u}}=\left( \cos
\theta ,\sin \theta \right) $. In particular, the combination of self-propulsion and diffusive reorientation allows
one to define an \emph{active persistence length} $\ell = F_{p}/\left(m \xi D_r \right)$, which in the overdamped limit gives the distance over which
a (free) particle's motion is correlated \cite{Marchetti2016}.  Interparticle interactions are described by a
WCA potential \cite{Weeks1971}
\begin{equation}
\Vwca(r) =
\left\{
\begin{array}{cc}
4 \epsilon \left[ \left( \frac{\sigma }{r}\right) ^{12}-\left( \frac{\sigma }{r%
}\right) ^{6}\right] +1 & r \leq 2^{1/6} \sigma  \\
0 & r > 2^{1/6} \sigma%
\end{array}
\right., \label{WCA-eq}
\end{equation}
%where $\sigma $ is the particle diameter. Finally, $\mathbf{\eta }^{T}(t)$
%and $\eta ^{R}(t)$ are delta-correlated thermal noises, $\langle \eta
%(t)\eta (t^{\prime })\rangle =\delta (t-t^{\prime })$ with corresponding
%diffusion coefficients $D$ and $D_{r} = 3D$.

In simulations we non-dimensionalize
using $\sigma $ as the unit length and $\tau =\sigma ^{2}/D$ as the unit
time, and we set the WCA well-depth parameter equal to the thermal energy, $\epsilon=\kt$. Denoting the new coordinates with primes, Eqs. \eqref{EOM-dimensional1-eq} - \eqref{EOM-dimensional3-eq} become
\begin{eqnarray}
\frac{d\mathbf{r}^{\prime }}{dt^{\prime }} &=&\mathbf{v}^{\prime } \label{EOM-dimensionless1-eq} \\
\frac{d\mathbf{v}^{\prime }}{dt^{\prime }} &=&3\xi_0 \ell _{0}
\widehat{\mathbf{u}}-\xi_0\nabla \Vwca-\xi_0 \mathbf{v
}^{\prime }+\sqrt{2}\xi_0\mathbf{\eta }^{T}(t^{\prime })  \label{EOM-dimensionless2-eq} \\
\frac{d\theta }{dt^{\prime }} &=&\sqrt{6}\eta ^{R}(t^{\prime }).  \label{EOM-dimensionless3-eq}
\end{eqnarray}

The parameter space is two-dimensional, spanned by the non-dimensional friction constant $
\xi_0= \xi \tau $ and active persistence length $\ell_{0}=\ell / \sigma$.

%The parameter space is two-dimensional, spanned by the non-dimensional friction constant, which gives the ratio of the diffusive timescale $\tau$ to the inertial timescale $\taui=\xi/m$: $
%\xi_0=\tau/\taui$,  and the active persistence length $\ell_{0}=\ell / \sigma$.
%In the rest of this paper we use equations \eqref{EOM-dimensionless1-eq} - \eqref{EOM-dimensionless3-eq}
%and drop the primes.

\begin{figure}
  \includegraphics[keepaspectratio=true,scale=0.4]{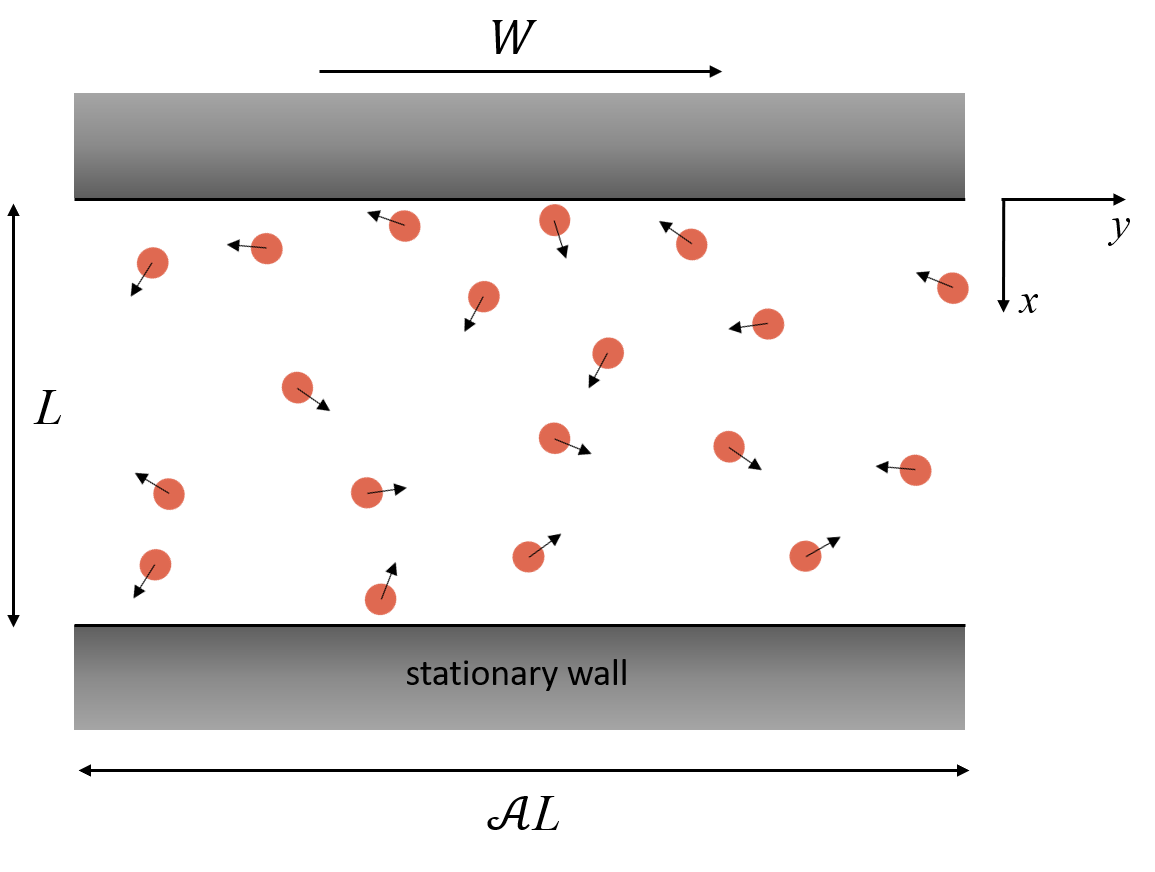}
  \caption{The shearing geometry. The aspect ratio $\mathcal{A}$ and packing fraction $\phi$ are varied to achieve a fixed overall number of particles.}
  \label{fig:shearing-geometry}
\end{figure}

\emph{Shearing geometry.} To understand the effects of boundary driving on this model, we consider a
simple shearing geometry (Fig. \ref{fig:shearing-geometry}), with periodic boundary conditions in the $y$ direction and confining walls in the $x$ direction. The bottom wall is stationary, and the upper
wall moves with constant velocity $W$. The $x$-component of the
particle-wall interactions is given by $\Vwca(x+\sigma)$ (equation \eqref{WCA-eq}), so that particles feel the wall potential for $x \leq (2^{1/6} - 1) \sigma \simeq 0.12 \sigma$.
In the $y$-direction a force $F_{w}$ drives the particles in the direction
of the wall's motion. We use a linear force,
$F_{w}=F_{w,0}\left( W-v_{y}\right)$, for a particle with velocity $%
v_{y}$. Unless noted otherwise, we take $F_{w,0} = 50$ and $W=5$ (in non-dimensional units).

Throughout the paper we will be calculating the flow velocity, $\langle v_{y}\rangle$, which is the average of $v_y$ over all $\mathbf{v}$ and $\theta$ at a given $\mathbf{r}$. Anticipating this, we define the average particle velocity `at the wall', $v_w$, to be the average velocity at $x \approx 0.35 \sigma$, i.e. slightly outside the range of the wall potential.
Note that in general $v_w \neq W$.

\emph{Simulation parameters.} Since we are interested in a range of values for the friction $\xi $, and
particularly the large friction limit, we use the Brownian dynamics
algorithm due to van Gunsteren and Berendsen, which is not limited by the
restriction $\Delta t\xi \ll 1$ \cite{vanGunsteren1982}. We set $\Delta t=0.00005 \tau$, and for a given $L$ and $\phi$ adjust the aspect ratio $\mathcal{A}$ (Fig. \ref{fig:shearing-geometry}) to give $10^4$ particles.  To rule out finite-size effects, we choose $L$  such that the channel dimensions are larger than any microscopic correlation length.  Since we only consider  values of $\ell$ and $\phi$ below the known onset of critical behavior and phase separation \cite{Redner2013,Marchetti2016}, the only correlation lengths to consider are those of a single particle trajectory in the absence of interactions, namely, $\ell$ and $\sqrt{D/ \xi}$. The former is the active persistence length, while the latter is the distance a particle with characteristic velocity $\sqrt{D \xi}$ travels in a frictional time $1/\xi$. Depending on the value of the friction,  $L$ in the range $25 \sigma$ - $100 \sigma$ is large enough to rule out finite-size effects due to these lengths (see appendix for details).  We begin recording statistics at $t=200\tau$, when all trajectories reach steady state, and  continue until $t = 1000 \tau$.

%In some cases we run several simulations in parallel in order to obtain improved statistics in sensitive cases, or to estimate errors  the sample variance \mfh{specify which ones somewhere}.

\section{Results} \label{sec-results}

In passive fluids described by the Boltzmann equation, the interaction
timescales are the smallest in the model, and therefore the primary
mechanism behind thermalization and relaxation into local equilibrium. In the case of a passive fluid interacting with a substrate, however, there exists an additional time scale, the frictional time $1/\xi$, which can be comparable to or smaller than the mean free time between collisions. Moreover, even in the absence of interparticle interactions, all momentum is dissipated into the substrate via the frictional mechanism. In these circumstances, momentum transport and dissipation are predominantly determined by the frictional and diffusive relaxation mechanisms, with interparticle interactions playing a supplementary role. Further, in the case of an active fluid, there exists the reorientation time $1/D_r$. This time scale influences how far the momentum from the wall penetrates into the bulk before being dissipated to the substrate. In the limit where the frictional and reorientation times are shorter than the mean free time, the phenomenology is most clearly understood by considering a system of non-interacting particles. Then, the non-interacting case can be used as a baseline to interpret the phenomenology when the interactions modify it. This is the route we follow below.

%In the active system, however, there are two additional relaxation timescales: the frictional time $1/\xi $ and the diffusive reorientation
%time $1/D_{r}$. These timescales are frequently comparable to, or
%smaller than, the mean free time between collisions. Moreover, even in the
%absence of interparticle interactions all momentum is dissipated into the
%substrate via the frictional mechanism. In these circumstances
%momentum transport and dissipation are predominantly determined by the
%frictional and diffusive relaxation mechanisms, with interparticle
%interactions playing a supplementary role. Thus, we begin our analysis
%with the non-interacting case to establish a baseline for understanding the interacting system.

\subsection{Dilute limit} \label{sec-dilute-limit}

%See, for instance, figure XXX
%with $\xi =10$ and varying $\ell _{p}$.

%\begin{equation}
%\langle v_{y}\rangle \approx e^{-\xi \langle t(x,u)\rangle },
%\end{equation}
%where $\langle t(x,u)\rangle $ is the average time for a particle to
%reach $x$ starting from $x=0$ with $v_{x}=u$. For passive particles, we
%can approximate $\langle t(x,u)\rangle \approx x/\sqrt{k_{B}T}=x/\sqrt{%
%D\xi }.$ So,%
%\begin{equation}
%\langle v_{y}\rangle \approx \exp \left[ -\frac{x\sqrt{\xi }}{\sqrt{D}}\right].
%\label{eq:ideal-flow-profile-passive}
%\end{equation}
%Physically, the decay length $\sqrt{D/\xi}$ is the distance a particle with characteristic velocity $\sqrt{k_B T}$ travels in a frictional time $1/\xi$: this is how far a particle penetrates into the bulk before its momentum is lost to the substrate.

%We consider the dilute (non-interacting) limit, using $
%W=5$ and $F_{w,0}=50$ in the dimensionless units. In general,

\begin{figure}
  \includegraphics[keepaspectratio=true,scale=0.23]{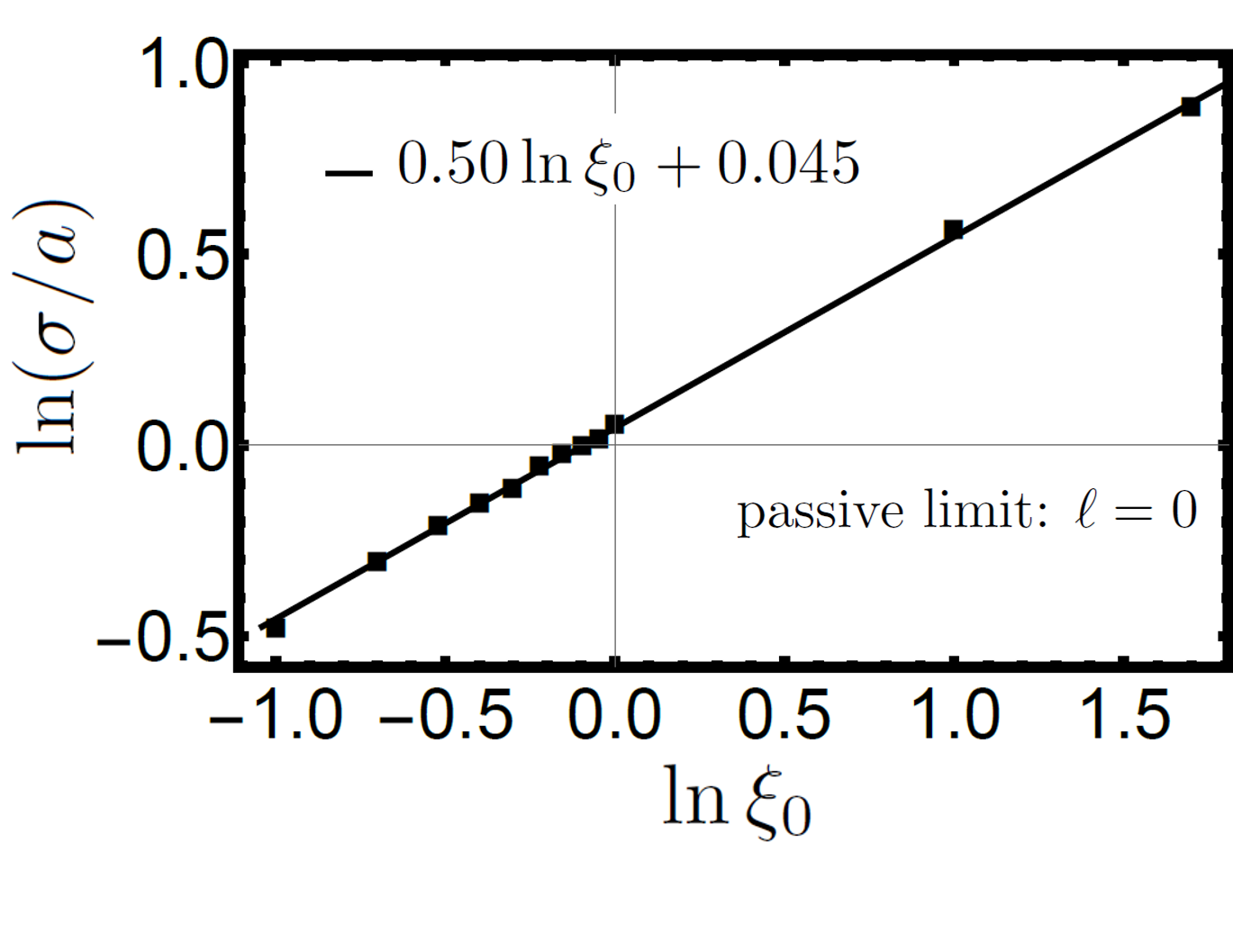}
  \includegraphics[keepaspectratio=true,scale=0.23]{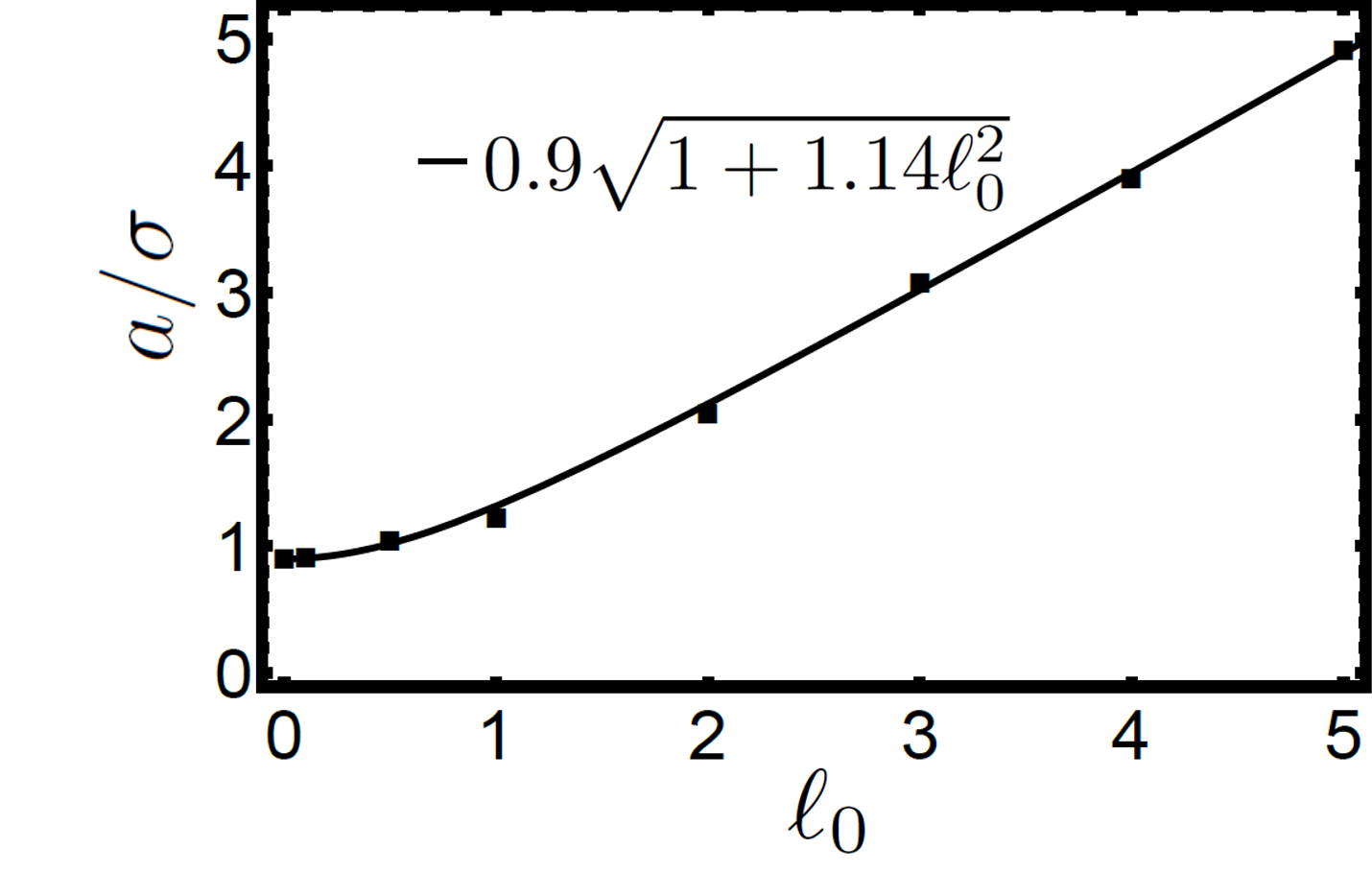}
  \caption{ Tests of the predictions in section \ref{sec-dilute-limit} for the dependence of the momentum transfer length scale $a$ on friction and activity. The value of $a$ at each parameter set is obtained by fitting the flow velocity profiles of non-interacting systems to the form $\langle v_y \rangle = e^{-x/a}$. The top panel shows the decay length as a function of  the nondimensional friction parameter $\xi_0$ in passive systems ($\ell_0=0$). The bottom panel shows $a$ as a function of active persistence length $\ell_0$ for fixed friction parameter $\xi_0=1$.  When fitting the value of $a$, we have excluded a boundary layer that exhibits deviations from exponential decay (see Appendix A and Fig.~\ref{fig:sample_fit}).}
    \label{fig:passive-active-fits}
\end{figure}

In the dilute limit, the steady-state flow velocity
$\langle v_{y}\rangle$ decays rapidly away from the boundary over a length
scale determined by $\xi $ and $\ell$. This trend reflects the fact that, by virtue of their persistent motion, particles travel
a short distance into the bulk before their momentum acquired at the
boundary dissipates into the substrate. The following kinetic picture can be
used to estimate the decay length. We start with an estimate for $\langle
v_{y}\rangle$:
\begin{equation}
\langle v_{y}\rangle \approx \exp \left[ -\frac{\xi}{\langle v_x \rangle} x\right],
\label{eq-ideal_vy_estimate}
\end{equation}
where $\langle v_x \rangle$ is the $x$-component of an appropriate characteristic velocity. In general, $\langle v_x \rangle$ is a function of $\xi$ and $\ell$. Thus, we define the decay length $a(\xi, \ell) \cong \langle v_x \rangle / \xi$, so that $\langle v_{y}\rangle \approx e^{-x/a}$.
%\begin{equation}
%\langle v_{y}\rangle \approx e^{-x/a}.
%\label{eq-ideal_vy_estimate-a}
%\end{equation}
For passive particles, there is only one possible microscopic length for $a$, coming from $\langle v_x \rangle =  \sqrt{D \xi}$. In this case
\begin{equation}
\langle v_{y}\rangle \approx \exp \left[ -\frac{1}{c} \frac{\sqrt{\xi }}{\sqrt{D}} x\right],
\label{eq:ideal-flow-profile-passive}
\end{equation}
where $c \approx 0.9$ is a numerical constant obtained from simulations. Physically, the decay length $a(\xi, \ell = 0) = c \sqrt{ D/\xi}$ is the distance a passive particle with characteristic velocity $\sqrt{D \xi}$ travels in a frictional time $1/\xi$: this is how far a particle penetrates into the bulk before its momentum is lost to the substrate.

With activity, a second characteristic velocity $v_a$ is introduced. An estimate for $\langle v_x \rangle$ can then be obtained as a root-mean-square combination of this velocity and $\sqrt{D \xi}$. This leads to a decay length $a(\xi, \ell)  = (c / \xi) \sqrt{D  \xi + \lambda v_a^2}$ where $\lambda$ is a fitting parameter. In general, $v_a$ will consist of $\ell_p$ times some characteristic time. For instance, if we assume that $\xi \gtrsim D_{r}=\mathcal{O}(1)$, i.e. frictional relaxation is faster that orientational relaxation, then the relevant timescale is $1/\xi$, and $v_a$ can be estimated as $F_{p}/(m\xi) = D_r \ell_p$. In the opposite limit, where orientational relaxation dominates, we have $v_a \sim F_p / (m D_r) = \xi \ell_p$. 
%\begin{equation}
%\langle v_x \rangle = a_1 \sqrt{k_B T + \left(a_2 D_r \ell\right)^2},
%\end{equation}
%so that
%\begin{equation}
%\langle v_{y}\rangle \approx \exp \left[ -\frac{x}{\ell _{p}}\right].
%\label{eq:ideal-flow-profile-active}
%\end{equation}%

We test these predictions by fitting them to the $\langle v_{y}\rangle$
profiles obtained from our simulations with non-interacting ABPs. As shown in Fig.~\ref{fig:passive-active-fits}, the results match the predicted
scaling well, provided a boundary layer which deviates from exponential
decay is excluded from the fit. A finer analysis which captures this part of the solution requires a full spectral analysis of the Fokker-Planck equation associated with Eqs. \eqref{EOM-dimensionless1-eq} - \eqref{EOM-dimensionless3-eq}, which we do not attempt here (see Ref. \cite{Wagner2017} for such an analysis in the context of a simpler model). We note that a similar asymptotic exponential decay has been found in steady-state \emph{density} profiles in active systems \cite{Enculescu2011,Elgeti2013,Lee2013,Yan2015,Duzgun2018}. In particular, Yan and Brady \cite{Yan2015} have found that knowledge of this part of the solution is sufficient for understanding a range of properties of the steady state.

%\begin{figure}
%  \includegraphics[keepaspectratio=true,scale=0.27]{passive.png}
%  \includegraphics[keepaspectratio=true,scale=0.25]{active.png}
%  \caption{ Tests of the predictions in section \ref{sec-dilute-limit} for the dependence of the momentum transfer length scale $a$ on friction and activity. The value of $a$ at each parameter set is obtained by fitting the flow velocity profiles of non-interacting systems to the form $\langle v_y \rangle = e^{-x/a}$ (see Fig.~\ref{fig:sample_fit}). The top panel shows the decay length as a function of  the nondimensional friction parameter $\xi_0$ in passive systems ($\ell_0=0$). The bottom panel shows $a$ as a function of active persistence length $\ell_0$ for fixed friction parameter $\xi_0=1$.  When fitting the value of $a$, we have discarded a boundary layer that exhibits deviations from exponential decay (see Appendix A and Fig.~\ref{fig:sample_fit}).}
%    \label{fig:passive-active-fits}
%\end{figure}

\subsection{Role of interactions} \label{sec-role-of-interactions}

%In passive fluids described by the Boltzmann equation, the interaction
%timescales are the smallest in the model, and therefore the primary
%mechanism behind thermalization and relaxation into local equilibrium. In
%the active system, however, there are two additional relaxation timescales
%to consider: the frictional time $1/\xi $ and the diffusive reorientation
%time $1/D_{r}=1/3$. These timescales are frequently comparable to, or
%smaller than, the mean free time between collisions. Moreover, even in the
%absence of interparticle interactions all momentum is dissipated into the
%substrate via the frictional mechanism. Thus, it may useful to think of
%momentum transport and dissipation as predominantly determined by the
%frictional and diffusive relaxation mechanisms, with interparticle
%interactions playing only a supplementary role. These ideas are addressed
%more systematically in section \ref{sec-theory}.

\begin{figure}
  \includegraphics[width=0.7\linewidth,height=0.46\linewidth]{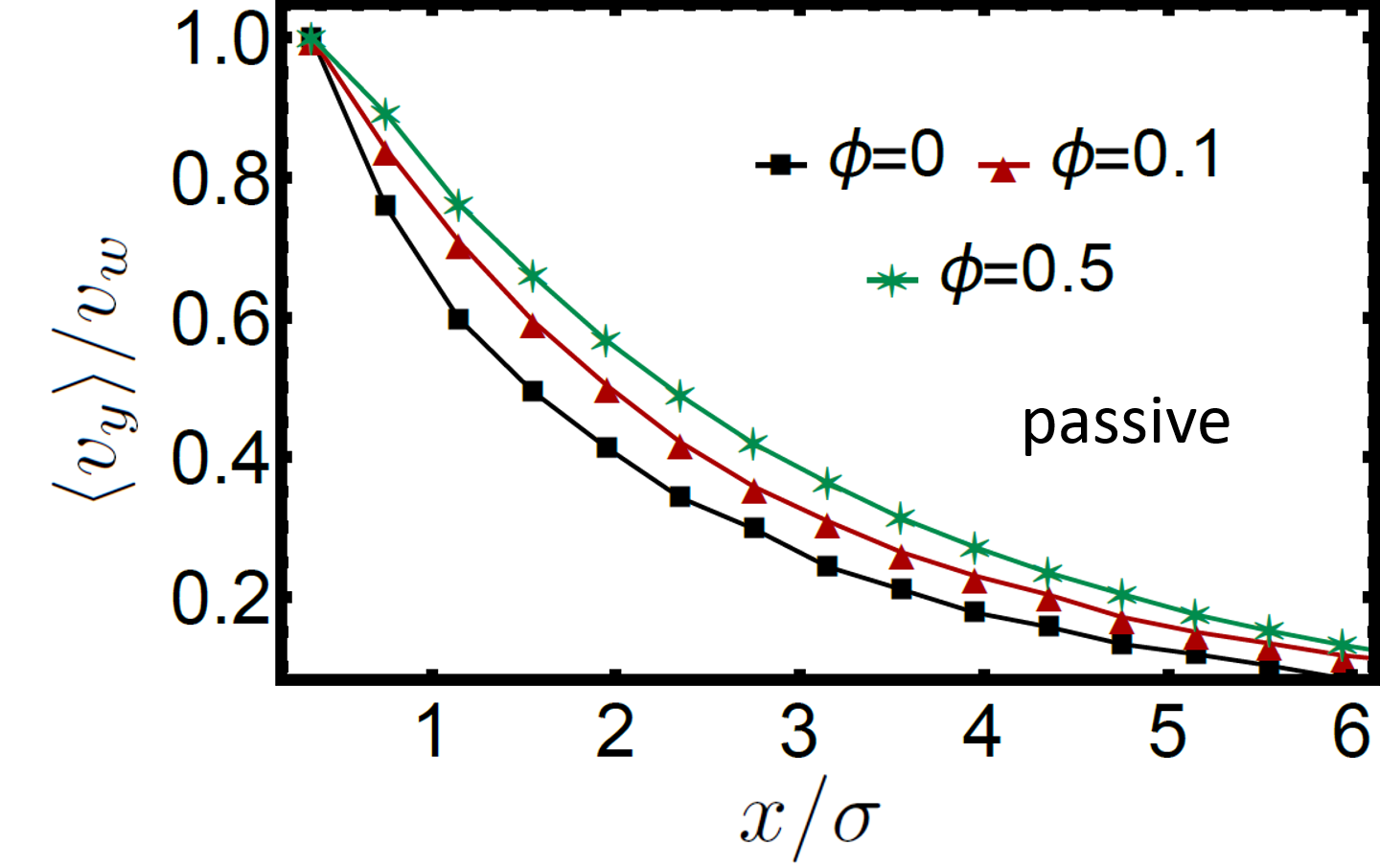}
  \includegraphics[width=0.7\linewidth,height=0.53\linewidth]{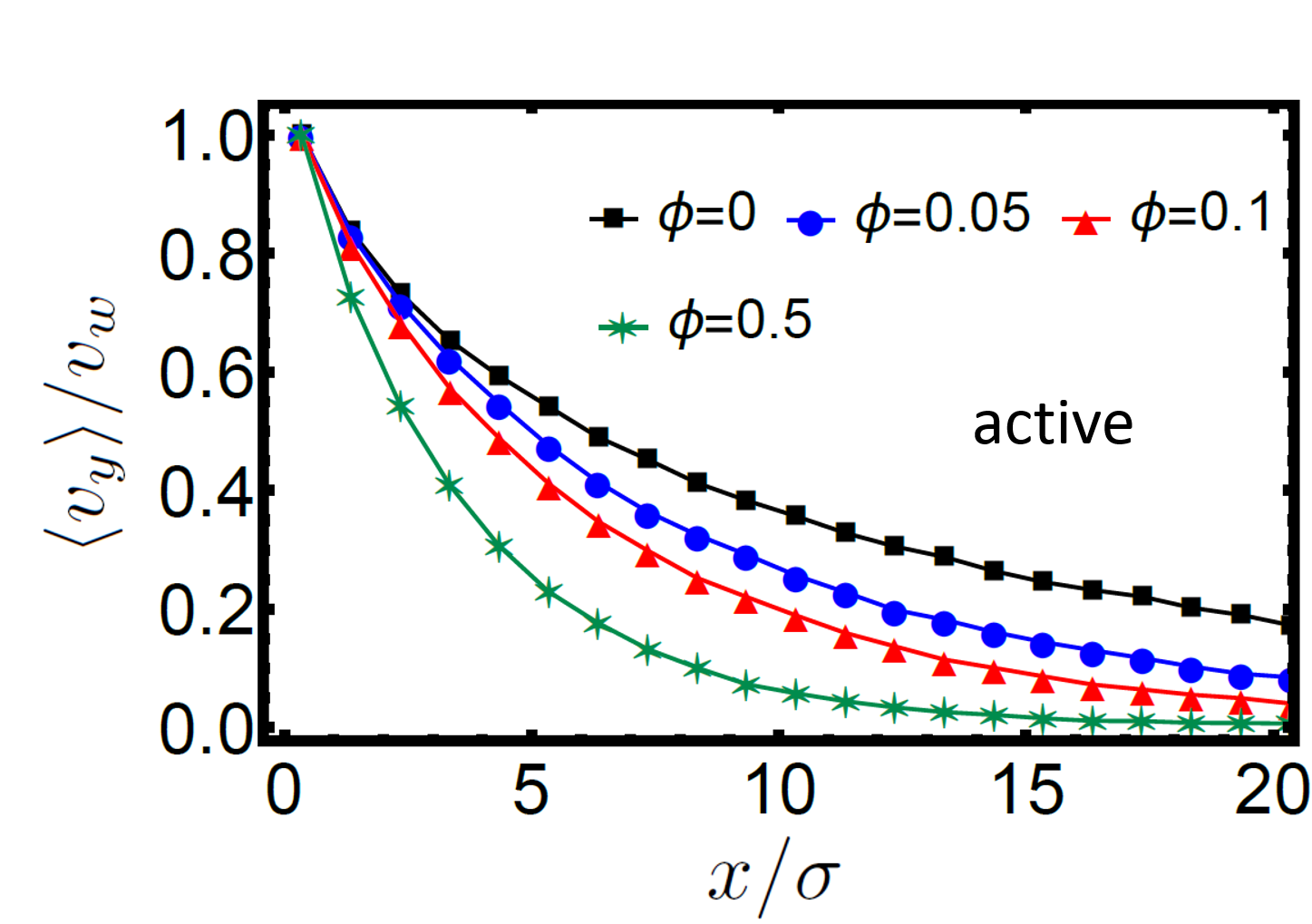}
  \caption{Illustration of the effect of interparticle interactions on momentum transport. The flow velocity $\langle v_y \rangle$ is shown as a function of distance $x$ from the boundary. For passive particles ($\ell_0 = 0$, top), increasing the packing fraction $\phi$ increases momentum transport, whereas with sufficiently high activity ($\ell_0 = 5$, bottom), the opposite is observed.}
  \label{fig:C1-C7}
\end{figure}

We now examine how interactions modulate the behavior of the non-interacting system. In general we find qualitatively similar
flow velocity profiles, with interactions either aiding or hindering momentum transport depending on the
values of $\xi_0$ and $\ell_0$. For instance, Fig. \ref{fig:C1-C7} shows the flow
velocity profiles for friction parameter $\xi_0 =0.1$ and several packing fractions for passive particles (Fig. \ref{fig:C1-C7} top) or active particles (Fig. \ref{fig:C1-C7} bottom). While increasing density increases momentum transport of passive particles, we observe the opposite effect  for the active case.

%This phenomenon can be explained in terms of a simple kinetic
%picture involving two density-dependent mechanisms of momentum transport. The first is the \emph{collisional transport} of momentum first considered by Enskog: at the instant of an interparticle collision, momentum is transferred across a length on the order of a particle diameter \cite{Chapman1970}. In the classical dilute gas, this mechanism results in a density-dependent increase to the viscosity. In our case, we represent this effect to first order in density as a correction, $j_{\text{coll}}$, to the overall momentum flux: $j_{\text{coll}} \propto v_{\text{avg}} \phi (d \langle v_y \rangle / dx)$, where $v_{\text{avg}}$ is the average particle speed in the dilute limit. 
%It is the contribution that remains if we set the particle diameter $\sigma$ to $0$.
This phenomenon can be explained with a more careful consideration of the density dependence of the total momentum flux $j_T$, defined as the flux of the $y$-component of momentum in the $x$-direction. We write $j_T$ as a sum of two contributions: a streaming contribution $j_{\text{stream}}$ and a collisional contribution $j_{\text{coll}}$. The streaming piece is the momentum flux due to the streaming motion of the particles between collisions. The collisional piece is the same as was first considered by Enskog: at the instant of an interparticle collision, momentum is transferred across a length on the order of a particle diameter \cite{Chapman1970}. In the classical dilute gas, this mechanism results in a density-dependent increase to the viscosity. If $v_{\text{avg}}(\rho(\mathbf{r}))$ is the (density-dependent) average particle speed, then we can write these contributions explicitly as

\begin{widetext}
\begin{align}
j_T &= j_{\text{stream}} + j_{\text{coll}} \\
 &\approx \underbrace{C_1 v_{\text{avg}}(\rho(\mathbf{r})) \cdot \underbrace{m \rho(\mathbf{r}) \langle v_y \rangle}_\text{momentum density}}_{j_{\text{stream}}} + \underbrace{C_2 v_{\text{avg}}(\rho(\mathbf{r})) \cdot \pi \sigma^2 \rho(\mathbf{r}) \cdot  m \rho(\mathbf{r}) \langle v_y \rangle + \mathcal{O}(\rho(\mathbf{r})^3)}_{j_{\text{coll}}} \label{eq:jt-exp}
\end{align}
\end{widetext}
where $C_1$ and $C_2$ are positive constants. As in the non-interacting case, the average speed $v_{\text{avg}}(\rho(\mathbf{r}))$ is estimated as a root-mean-square combination of the thermal velocity $\sqrt{D \xi}$ and a characteristic active velocity. For the latter quantity we previously used $v_a$, defined as the characteristic velocity due to free active motion in the absence of interparticle interactions. On the other hand, at finite density interparticle interactions tend to block this free motion, which results in an effective decrease in the characteristic active velocity \cite{Stenhammar2013,Hancock2017}. Let this new effective velocity be denoted by $v_a^I(\rho(\mathbf{r}))$, a function of $\rho$. At low densities, we can expand to first order in $\rho$, giving $v_a^I(\rho(\mathbf{r}))  = v_a \left[ 1 - C_3 \sigma^2 \rho(\mathbf{r}) + \mathcal{O}(\rho(\mathbf{r})^2) \right]$, where $C_3 > 0$. Putting these pieces together, we have
%
%
% is written with a density dependence since the average active velocity experiences an effective decrease at finite density due to increased collisional frequency with other particles \cite{Stenhammar2013}. At low densities, this effective active velocity can be expanded to first order in density:
%
%To make further progress, we expand $v_{\text{avg}}(\rho(\mathbf{r}))$ to first order in density:
%
\begin{align}
v_{\text{avg}}(\rho(\mathbf{r})) &=   \sqrt{v_a^2 \left[ 1 - C_3 \sigma^2 \rho(\mathbf{r}) + \mathcal{O}(\rho(\mathbf{r})^2) \right]^2  + D \xi} \\
&= v_a - C_3 \sigma^2 \frac{v_a^2}{v_0} \rho(\mathbf{r}) + \mathcal{O}(\rho(\mathbf{r})^2)
\end{align}
where $v_0 \equiv\sqrt{v_a^2  + D \xi}$. Substituting into \eqref{eq:jt-exp} gives

\begin{align}
j_T &= C_1 v_0 m \rho(\mathbf{r}) \langle v_y \rangle \\
&-  C_1 C_3 \sigma^2 \frac{v_a^2}{v_0}  m \rho(\mathbf{r})^2 \langle v_y \rangle \\
&+ C_2 v_0 m \rho(\mathbf{r})^2 \pi \sigma^2 \langle v_y \rangle + \mathcal{O}(\rho(\mathbf{r})^3).
\end{align}
Since we are interested in behavior in the mean, we make a further approximation and set $\rho(\mathbf{r}) = \frac{4}{\pi} \phi$, where $\phi$ is the overall packing fraction (independent of space). 

In steady state, conservation of the $y$-component of momentum says

\begin{align}
\frac{d j_T}{dx} &= (\text{momentum source/sink}) \\
&= -\xi m \rho(\mathbf{r})  \langle v_y \rangle \\
&\approx -\xi m (4 / \pi) \phi  \langle v_y \rangle
\end{align}
Substituting $j_T$ and absorbing numerical factors into the constants gives

\begin{align}
\frac{d \langle v_y \rangle}{dx} = -\xi \left\{ C_1 v_0 \left[1 + \left(\frac{C_2}{C_1} - C_3 \frac{v_a^2}{v_0^2} \right) \sigma^2 \phi \right] \right\}^{-1} \langle v_y \rangle.
\end{align}
Thus, the decay length of the flow velocity profile increases with $\phi$ if

\begin{equation}
\frac{C_2}{C_1} - C_3 \frac{v_a^2}{v_a^2  + D \xi} > 0
\end{equation}
which is certainly true if $v_a = 0$, since $C_1$ and $C_2$ are positive. With large enough activity, however, the quantity becomes negative, and the decay length decreases with $\phi$. The transition occurs when
\begin{align}
\frac{C_2}{C_1} - C_3 \frac{v_a^2}{v_a^2  + D \xi} = 0 \\
\rightarrow v_a = C  \sqrt{D \xi}
\label{eq:phase-boundary-generic}
\end{align}
where $C$ is a new constant. 

\begin{figure}
  \includegraphics[width=1\linewidth,height=0.7\linewidth]{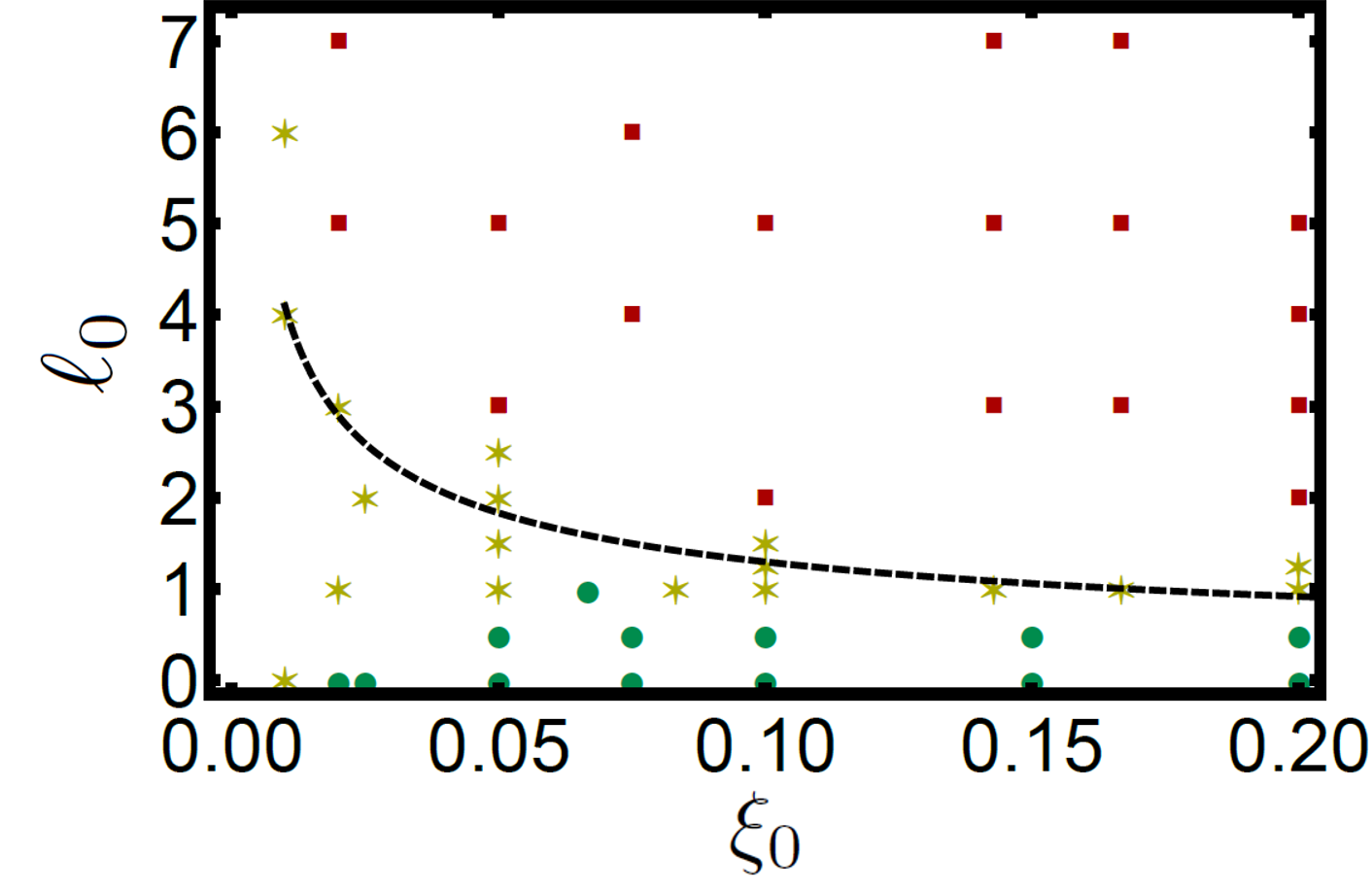}
    \includegraphics[width=1\linewidth,height=0.7\linewidth]{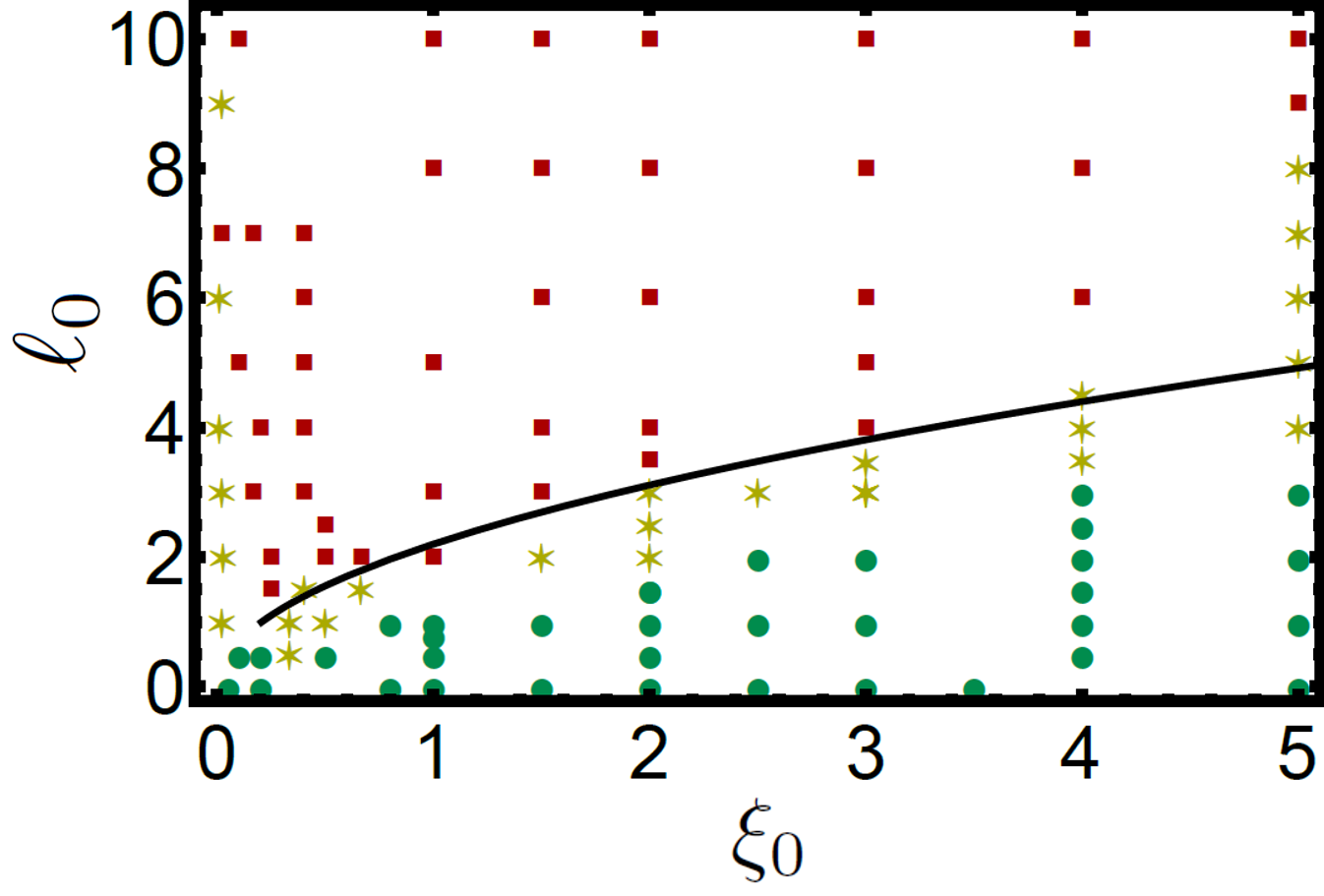}
  \caption{Generalization of figure \ref{fig:C1-C7} to the entire ($\xi_0$, $\ell_0$) space. The top and bottom panels respectively show the low-and high-friction regimes. Using the criterion discussed in the main text, red squares indicate parameter values for which interactions hinder transport, and green circles show values for which interactions aid transport. The solid and dashed curves are estimates for the boundary between the two regimes, based on the kinetic picture discussed in section \ref{sec-role-of-interactions}.}
     \label{fig:phase-diagram}
\end{figure}

In the previous section, we estimated $v_a \sim F_{p}/(m\xi) = D_r \ell_p$ in the limit $\xi \gg D_r$, and $v_a \sim F_p / (m D_r) = \xi \ell_p$ in the opposite limit. Using these estimates, we arrive at the following predictions for the boundary in the $(\ell_0, \xi_0)$ space:
%To complete the picture, we need to estimate $v_a$. If we assume that $\xi \gtrsim D_{r}=\mathcal{O}(1)$, i.e.
%frictional relaxation is faster that orientational relaxation, then $v_a$ can be estimated as $F_{p}/(m\xi) $. In the opposite limit, where orientational relaxation dominates, we have $v_a \sim F_p / (m D_r)$. This leads to the following predictions for the boundary in the $(\ell_0, \xi_0)$ space:

\begin{equation}
\ell_0 = a_1 \xi_0^{1/2}, \hspace{3mm} \xi_0 \gtrsim 1;
\label{eq:pred-1}
\end{equation}
and
\begin{equation}
\ell_0 = a_2 \xi_0^{-1/2}, \hspace{3mm} \xi_0 \lesssim 1,
\label{eq:pred-2}
\end{equation}
where $a_1 = 1.96$ and $a_2 = 0.28$ are constants obtained by fitting the expressions to a phase diagram in the $(\ell_0, \xi_0)$ space. The phase diagram is shown in Fig. \ref{fig:phase-diagram}, together with the fits (Eq. \eqref{eq:pred-1}, solid line; Eq. \eqref{eq:pred-2}, dashed line). Red squares denote systems where interactions hinder transport, green circles where interactions aid transport, and yellow stars where the result is indeterminate. We classify each simulation to one of these categories by comparing the penetration depth of the flow velocity profile (c.f. Fig. \ref{fig:C1-C7}) at $\phi = 0, 0.05, 0.1, 0.2$. (explained in more detail in appendix B). We obtained the prefactors in Eqs. \eqref{eq:pred-1} and \eqref{eq:pred-1} by performing a best fit on the yellow (indeterminate) data points in the ranges $0.5 \leq \xi_0 \leq 5$ and $0.02 \leq \xi_0 \leq 0.2$, for Eqs. \eqref{eq:pred-1} and \eqref{eq:pred-2} respectively.

An alternate argument to arrive at the same predictions proceeds by comparing the respective length scales over which momentum is transported due to (A) thermal motion and (B) free active motion. In the absence of interactions, the first length scale corresponds to free Brownian motion: independent of active driving, particles travel a distance $\sim \sqrt{D/\xi }$ before their initial $y$-momentum is dissipated into the substrate. The second length scale is determined by a characteristic active velocity $v_a$, which (again in the absence of interactions) causes particles to travel
an average distance of $v_a / \xi$ before losing their initial $y$-momentum. 

%Finally, length scale (C) is simply the particle diameter $\sigma$. 

Interparticle interactions affect transport over these two length scales differently. Note first that interactions do not interrupt transport over length scale (A), since linear momentum is conserved during the (nearly instantaneous) collisions. In fact, interactions slightly aid transport in this case since collisions also involve instantaneous and lossless transform of momentum over a particle diameter. On the other hand, particle \textit{orientation} is not transferred in collisions, i.e. a particle which would otherwise carry its $y$-momentum
over a length $v_a / \xi$ might transfer its momentum to a particle
oriented in the opposite direction, breaking transport across this length.
Thus, interactions interfere with transport over length scale (B). 

In light of these conclusions, it is reasonable to expect that in
cases where length scale (A) dominates (B), interactions aid momentum transport; whereas when (B)
dominates (A), the opposite is observed. The boundary between the two behaviors occurs when the length scales (A) and (B) are comparable: $v_a / \xi \sim \sqrt{D/\xi }$. This result agrees with Eq. \ref{eq:phase-boundary-generic}.

\subsection{Structure and transport at the wall}

The system exhibits further nontrivial behavior \emph{at} the wall (i.e. within roughly a particle diameter of the wall). Understanding the behavior in this region will be important for establishing proper boundary conditions on any continuum theory describing the bulk. 

\subsubsection{Flow reversal}

\begin{figure}
  \includegraphics[keepaspectratio=true,scale=0.3]{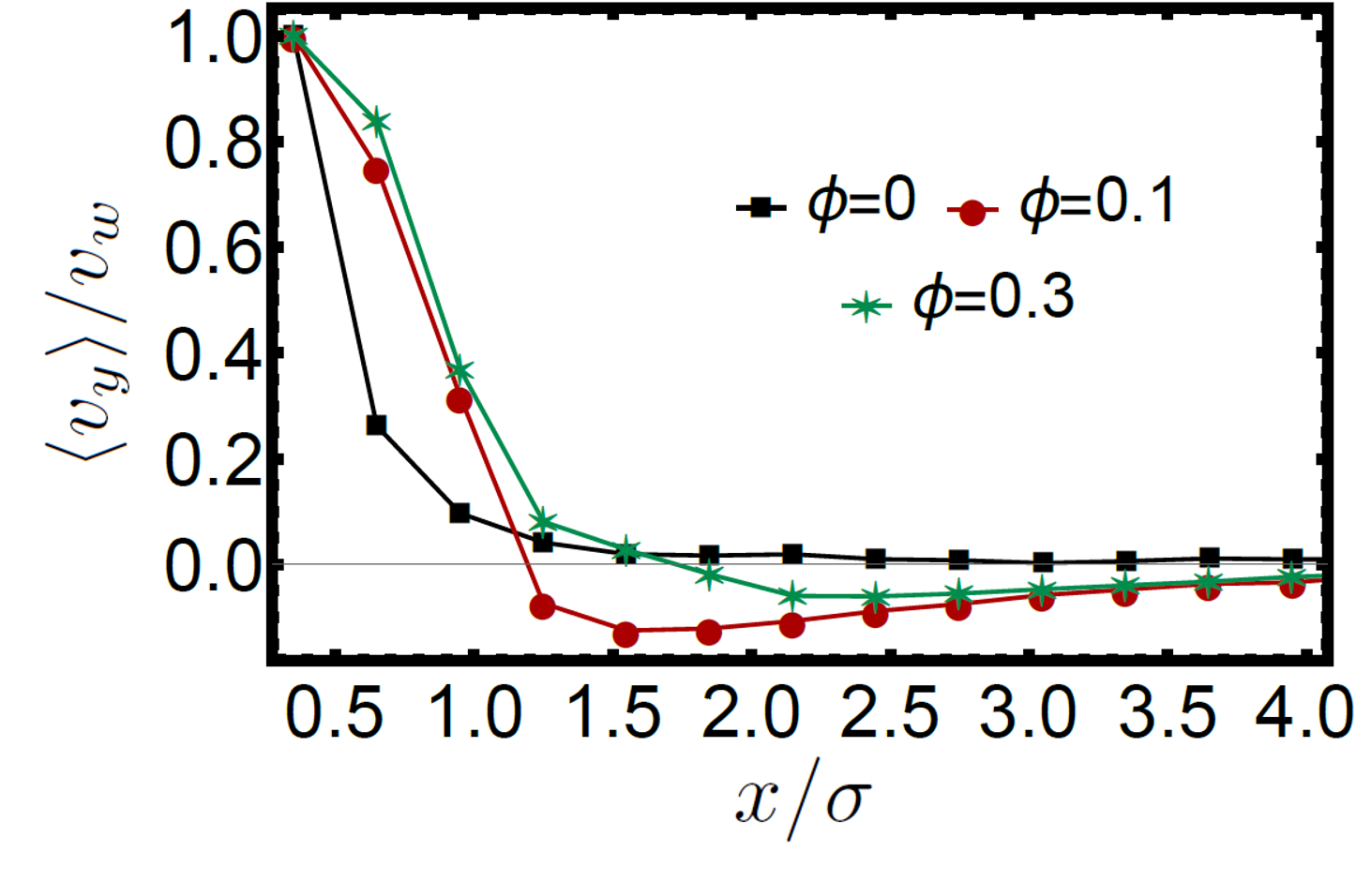}
  \includegraphics[keepaspectratio=true,scale=0.3]{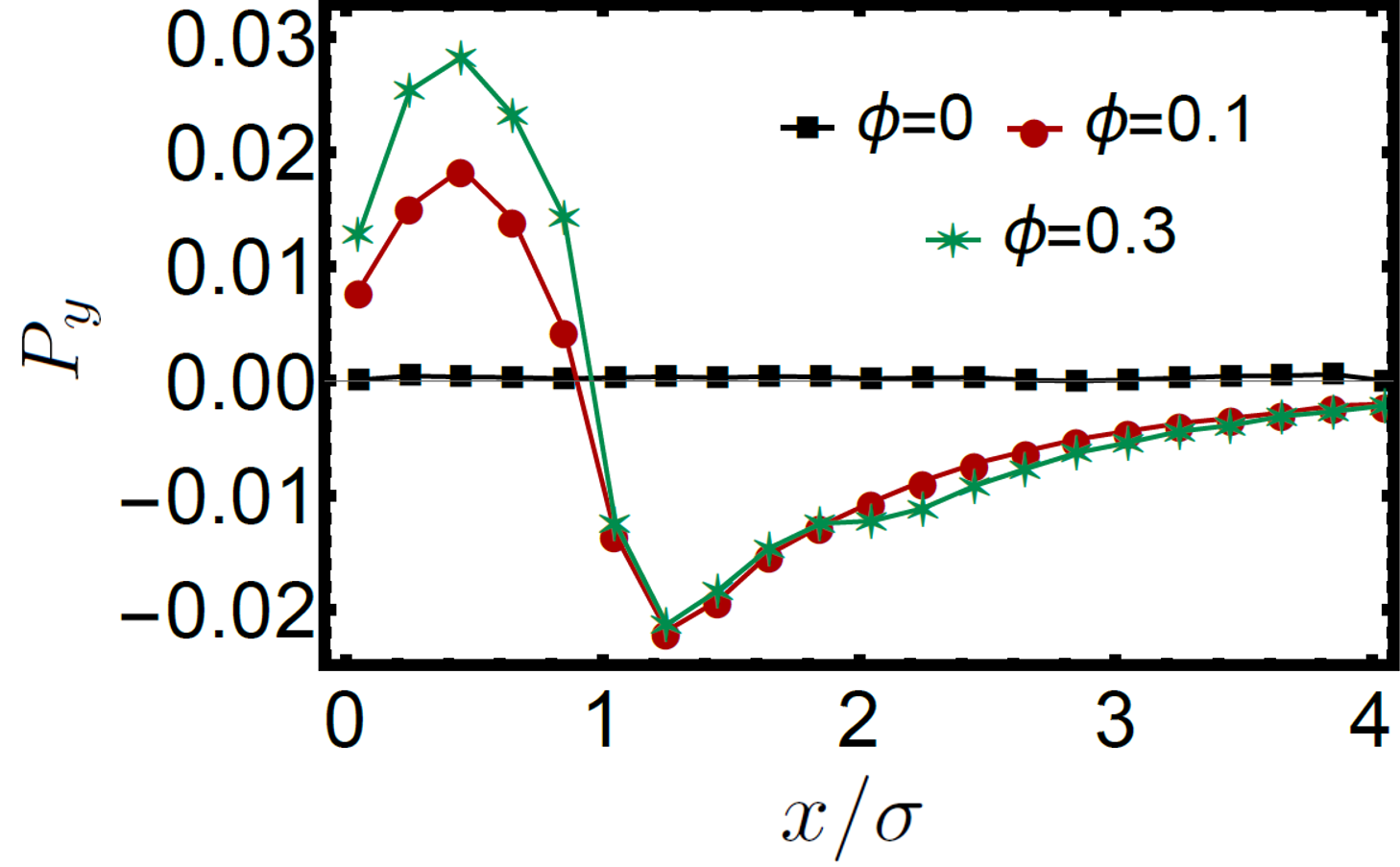}
  \caption{An instance of flow reversal. The top plot shows the flow velocity as a function of distance from the wall $x$, and the bottom plot shows the $y$-component of the polarization. The ``kinetic sorting" mechanism illustrated in Fig. \ref{fig:kinetic_sorting} induces this polarization, which itself generates the negative flow velocity seen in the top plot. The sorting mechanism depends on interparticle interactions and is optimized for intermediate values of the packing fraction $\phi$. The uncertainty on each data point is negligible compared to the symbol size. Other parameter values are $\xi_0=30$, $\ell_0=5$, and $W = 5$.}
  \label{fig:flow-reversal}
\end{figure}

 First, we observe \emph{flow reversal} near the boundary in some parameter ranges. For instance, Fig. \ref{fig:flow-reversal} shows the flow velocity profiles for $\xi_0=30$, $\ell_0=5$ and several packing fractions. In this case flow reversal occurs for the intermediate packing fraction: $\phi \approx 0.1$.

%\begin{cgw}Need to say something about vorticity/turbulence in classical fluids\end{cgw}
%The apparent paradox in this phenomenon is the existence of a nonzero flow in a direction opposite to any external driving. On the other hand, the system is driven at the particle level due to activity, and it is well known that this driving alone is sufficient
This flow reversal phenomena is reminiscent of other behaviors in active systems that would be thermodynamically forbidden at equilibrium, such as spontaneous flow \cite{Tailleur2009,Angelani2009,Ghosh2013} and orientational order in the absence of torques \cite{Enculescu2011,Wagner2017}. The operating principle underlying these phenomena is the ability of conservative fields and geometric confinement to kinetically ``sort" particles from an isotropic state into an orientationally ordered one. A similar mechanism drives flow reversal here, with interparticle interactions playing the role of the ``sorting" force. More precisely, an active system initialized with the $y$-component of the polarization $P_y$ equal to 0 (the steady state in the absence of activity) evolves towards a state with $P_y \neq 0$.

\begin{figure}
  \includegraphics[keepaspectratio=true,width=1.5\linewidth,height=0.3\linewidth]{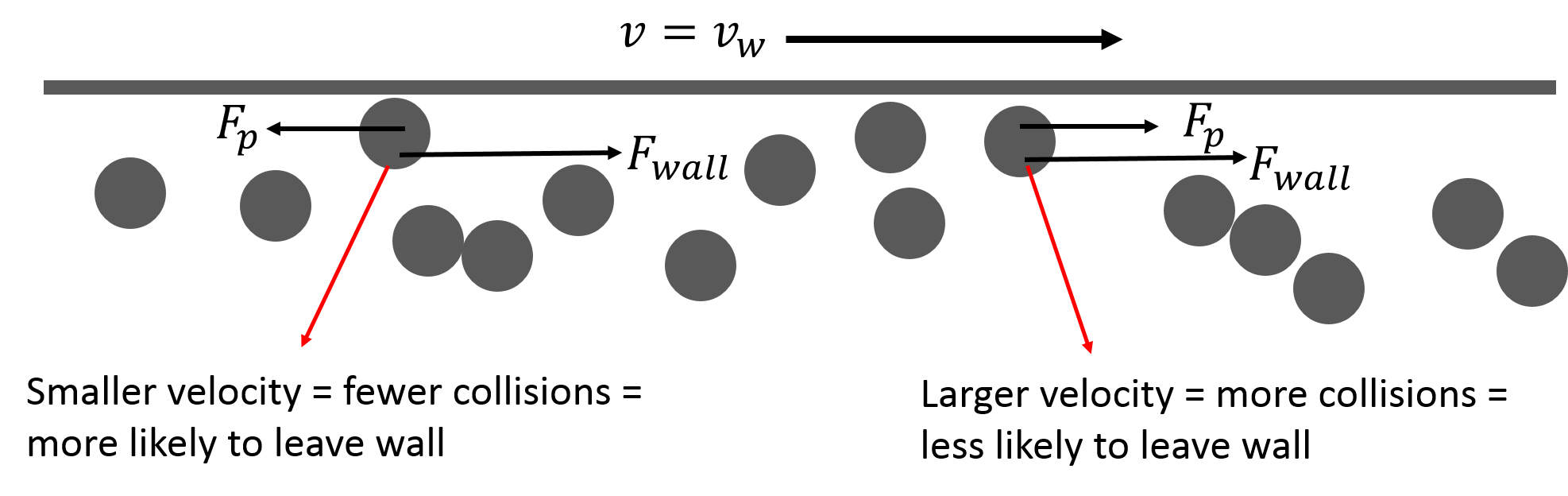}
  \caption{An illustration of the kinetic sorting mechanism discussed in the main text. Particles are split into two types: those with orientation parallel to the driving (type A), and those with orientation opposite the driving (type B). Compared with their type B counterparts, type A particles on average undergo more collisions with particles in the secondary layer (at $\sim 2\sigma$ from the wall). Since these collisions tend to push particles back towards the wall, the system sorts into a polarized steady state where there are more type A particles at distance $~\sigma$, and more type B particles at distance $2\sigma$.}
  \label{fig:kinetic_sorting}
\end{figure}

The mechanism is illustrated in Fig. \ref{fig:kinetic_sorting}. We divide particles near the wall into two layers at distances $\approx\sigma$ and $2 \sigma$ from the wall. We consider friction sufficiently large that the outer layer has a much smaller flow velocity than the inner layer.  Let us now consider two types of particles in the inner layer: type A oriented parallel to the direction of driving ($+y$-direction, $0 < \theta < \pi$), and type B antiparallel ($-y$-direction, $-\pi < \theta < 0$). Suppose $P_y = 0$. Then,
\begin{align}
|\langle v_{y}\rangle_A| &-  |\langle v_{y}\rangle_B| \approx |v_w +  \alpha D_r \ell| - |v_w - \beta D_r \ell|,
%&> \hspace{3mm} \left\{
%\begin{array}{cc}
%3 \left( \alpha +\beta \right) \ell _{p} & 3 \beta \ell _{p}<v_{w} \\
%2v_{w}+3 \left( \alpha -\beta \right) \ell _{p} & 3 \beta \ell _{p}\geq v_{w}%
%\end{array}
%\right.,
\end{align}
where $0 \leq \alpha, \beta \leq 1$ are constants.  In practice, $\alpha$, $\beta$ are close to $1$ since in the absence of boundary driving, orientations near the wall typically concentrate near $\pm \pi/2$. 

The variation of $|\langle v_{y}\rangle_A| -  |\langle v_{y}\rangle_B|$ with $\ell$ can be made clearer by rewriting the absolute values:

%In particular, $|\langle v_{y}\rangle_A| -  |\langle v_{y}\rangle_B|$ is always greater than $0$, as can be seen by rewriting the absolute values as
\begin{align}
|\langle v_{y}\rangle_A| &-  |\langle v_{y}\rangle_B|  =  \hspace{3mm} \left\{
\begin{array}{cc}
 \left( \alpha +\beta \right) D_r \ell & \beta D_r \ell <v_{w} \\
2v_{w}+ \left( \alpha -\beta \right) D_r \ell &  \beta D_r \ell \geq v_{w}
\end{array}
\right..
\end{align}

For $0 < \beta D_r \ell < v_w$, $|\langle v_{y}\rangle_A| -  |\langle v_{y}\rangle_B|$ is positive, increasing linearly with $\ell$ starting from $0$. On the range $\beta D_r \ell \geq v_{w}$, it either increases or decreases with $\ell$ depending on the sign of $\alpha - \beta$. In any case, however, since  $\alpha \approx \beta$, we can expect $|\langle v_{y}\rangle_A| -  |\langle v_{y}\rangle_B|$ to be positive over a large range of $\ell$. In what follows we therefore assume the positivity of $|\langle v_{y}\rangle_A| -  |\langle v_{y}\rangle_B|$.

%For simplicity we assume $|\langle v_{y}\rangle_A| -  |\langle v_{y}\rangle_B|$ is positive  (this is likely true for a large range of $\ell$, since $\alpha \approx \beta$). 

Now, the crux of the argument is this: Since particles of type A on average possess velocities larger in magnitude than those of type B, they undergo more off-center collisions with particles in the outer layer. Since these types of collisions tend to push particles back towards the wall, type B particles can more easily escape the inner layer. In other words, if $r_X^{I \rightarrow O}$ is the rate of particle species $X$ traveling from the inner to outer layer, then $r_B^{I \rightarrow O} > r_A^{I \rightarrow O}$. On the other hand, this asymmetry is not as pronounced for particles traveling from the outer layer to the inner one: $r_B^{O \rightarrow I} \approx r_A^{O \rightarrow I}$ (since the difference between A and B velocities is smaller in the outer layer). The overall imbalance of rates implies that the $P_y = 0$  state is not stable. We verify this prediction with simulation results, observing $P_y > 0$ in the inner layer and $P_y < 0$ in the outer layer (Fig. \ref{fig:flow-reversal}). Finally, orientational order can be connected to the flow velocity  if $\xi \gg D_r$ and $\ell > \sqrt{D/\xi}$, i.e. active driving dominates thermal driving. In this case $\mathbf{v}$ is approximately parallel to $\widehat{\mathbf{u}}$, and therefore $P_y < 0$ corresponds to a negative contribution to the overall flow velocity.
%the rate of type A particles transitioning from the outer to inner layer is roughly equal to the corresponding rate for type B particles.
%at distance $~\sigma$ of the wall

\begin{figure}
  \includegraphics[keepaspectratio=true,width=0.9\linewidth]{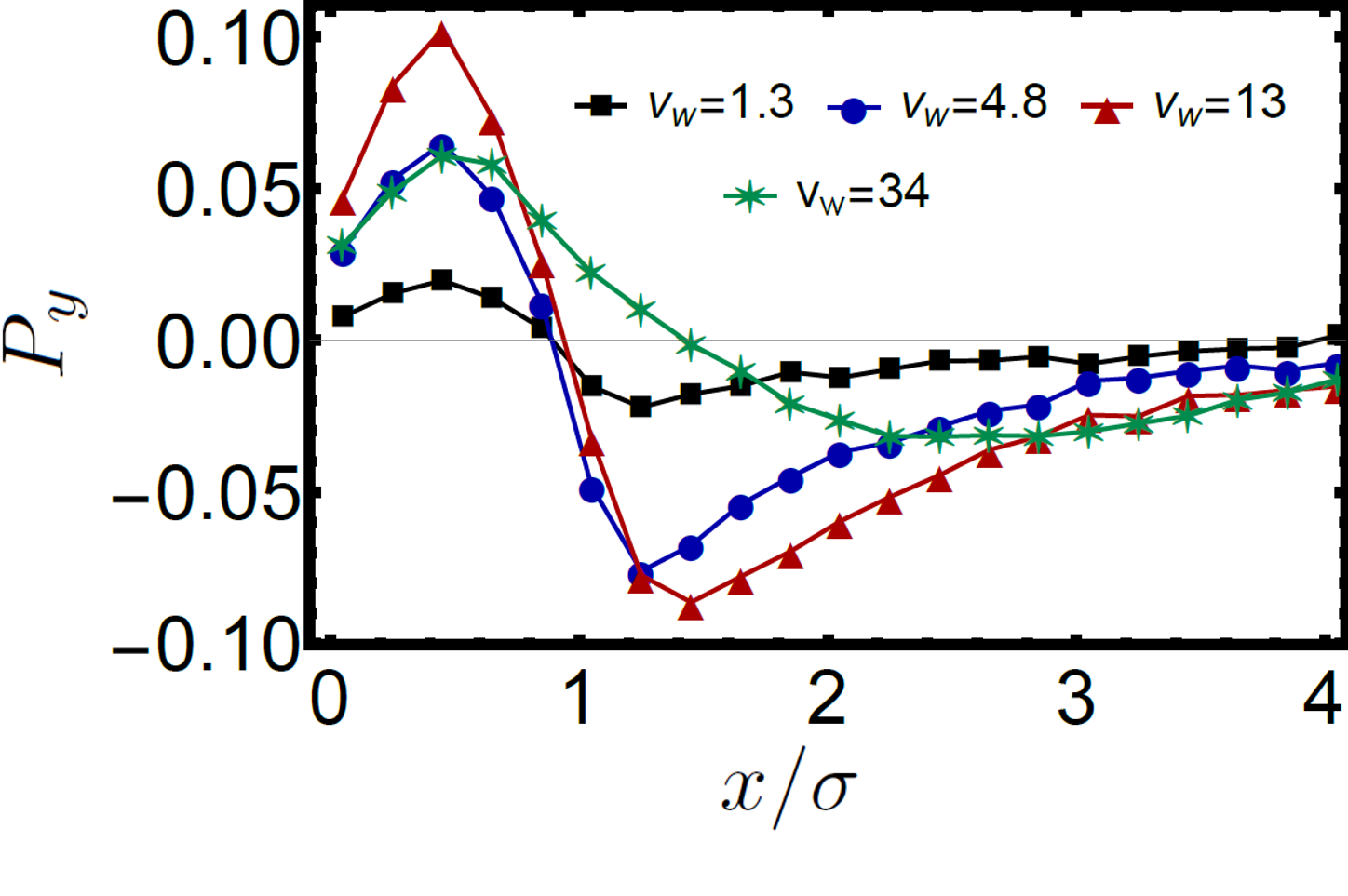}
  \includegraphics[keepaspectratio=true,width=0.9\linewidth]{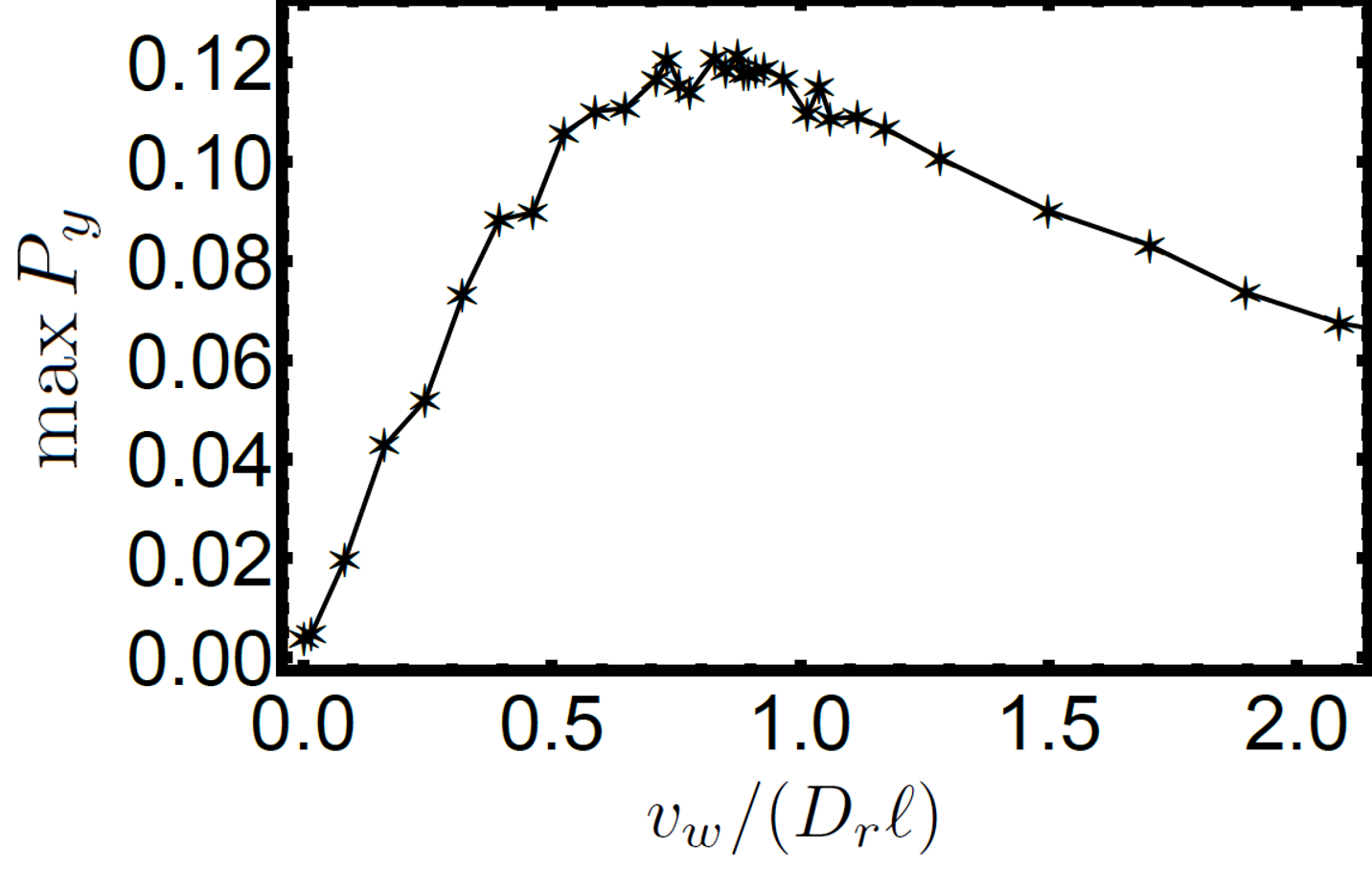}
  \caption{Dependence of the flow reversal phenomenon on driving. The top panel shows the $y$ polarization for varying values of $v_w$ (the particle velocity at the wall). In simulations $v_w$ is varied indirectly varied by changing the parameter $W$ in the wall force $F_{w}=F_{w,0}\left( W-v_{y}\right)$, over the range $W\tau^2/m\sigma\in[0.1, 200]$. The bottom panel shows the \emph{maximum} value of $P_y$ for each $v_w$. The magnitude of the induced polarization is maximized when $D_r \ell \sim v_{w}$. Here $\xi_0=30$, $\ell_0=5$, and $\phi = 0.1$.}
  \label{fig:resonance}
\end{figure}

In fact, it is possible to make a more precise prediction. Since $|\langle v_{y}\rangle_A| -  |\langle v_{y}\rangle_B|$ increases monotonically with $v_w$ until $D_r \ell \sim v_{w}$, we expect the induced polarization to also increase with $v_w$ until $D_r \ell \sim v_{w}$. This trend is confirmed in Fig. \ref{fig:resonance}. On the other hand, further increasing $v_w$ results in a smaller polarization, despite the fact that $|\langle v_{y}\rangle_A| -  |\langle v_{y}\rangle_B|$ saturates. This suggests that for large driving, particles of both types A and B collide so frequently that the asymmetry between the two is washed out, i.e. the difference in rates $r_B^{I \rightarrow O} - r_A^{I \rightarrow O}$ is no longer proportional to $|\langle v_{y}\rangle_A| -  |\langle v_{y}\rangle_B|$.

%This suggests that more complex mechanisms come into play for large driving.

%Repeating the same simulations while fixing $\phi$ and varying other parameters reveals that the phenomenon occurs only for sufficiently large friction and activity, and for intermediate driving ($v_w$ on the order of the terminal active velocity $3 \ell$).
%
%Besides this fact, a particle's velocity decorrelates with its orientation as $\xi$ becomes small.

\subsubsection{Stress at the boundary}

%\begin{cgw}Say something about generic stress-strain relation, generalized notion of viscosity, and the fact that viscosity is less than useful for this system since e.g. it depends in a complex fashion on boundary conditions and space\end{cgw}
%Despite the complexity of the stress-strain response in the bulk, it is illuminating to look at the stress at the wall, $\sigma_w$.
%(Here stress is [average force @ wall]/[unit length]).

From an experimental standpoint, an important observable is the stress at the wall, $\sigma_w$, defined as the average force at the wall per unit length. Note that since the wall potential \eqref{WCA-eq} has finite width in our simulations, our measurement of wall stress  includes all particles with $x/\sigma < 2^{1/6} - 1 \approx 0.123$.

\begin{figure}
  \includegraphics[keepaspectratio=true,width=0.43\linewidth]{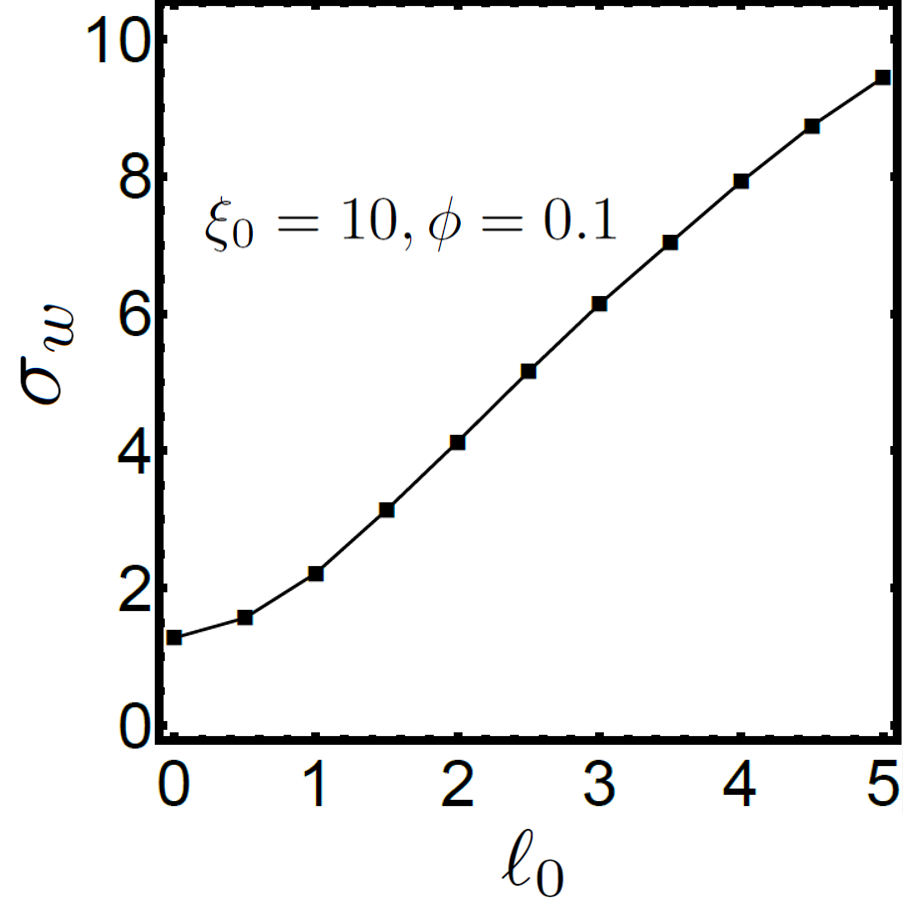}
  \includegraphics[keepaspectratio=true,width=0.45\linewidth]{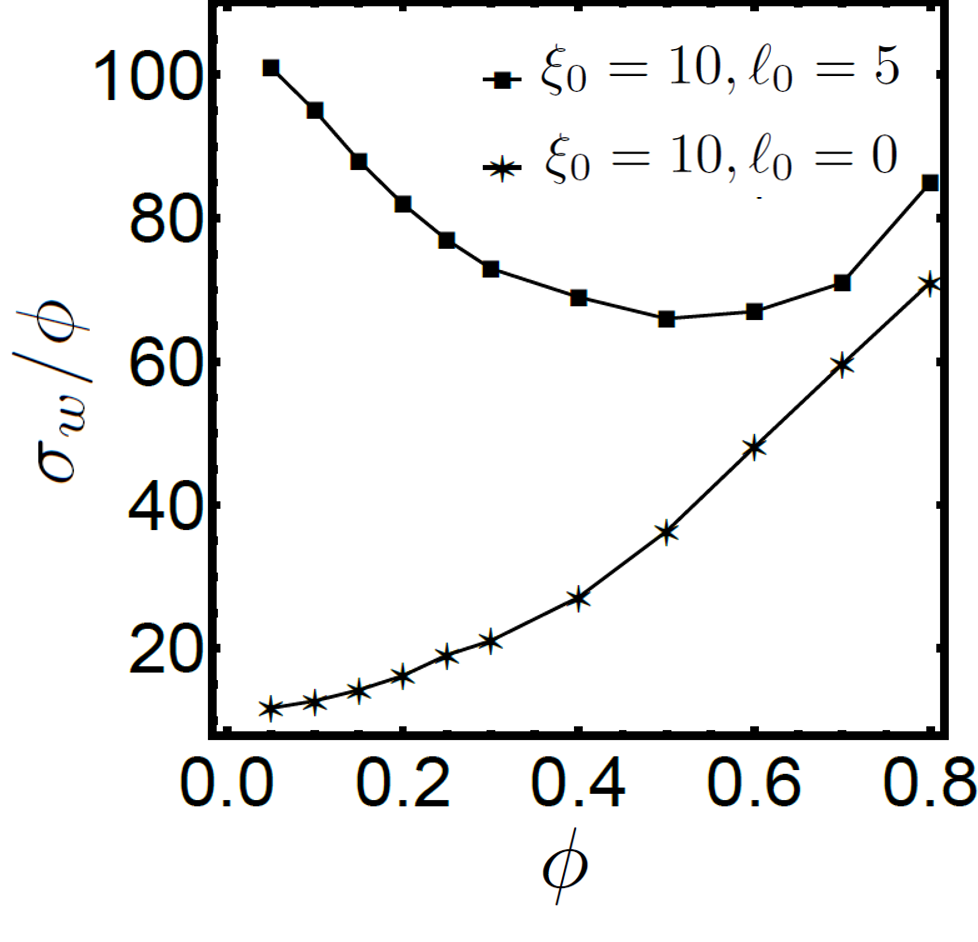}
  \caption{Left: Stress at the wall $\sigma_w$ as a function of $\ell_0$.The stress increases monotonically, corresponding to increasing particle accumulation at the boundary. Right: $\sigma_w$ as a function of packing fraction $\phi$. To illustrate more clearly the effect of interactions, the $y$-axis shows $\sigma_w / \phi$: this is to factor out the linear dependence $\sigma_w \propto \phi$ already present in the non-interacting system.}
  \label{fig:stress-strain}
\end{figure}

We observe that  $\sigma_w$ increases monotonically with activity (\ref{fig:stress-strain}). This result can be explained by the fact that more particles accumulate at the wall with higher activity, increasing the burden of shearing the system.
%On the other hand, The
%second figure shows a rapid jump in stress between l\_p = 15 and l\_p = 20.
%I have to check this visually, but I'm pretty sure this is happening due to
%phase separation.
However, we observe more complex behavior with varying density. First, we note that in a non-interacting system the stress
increases linearly with packing fraction $v_w \thicksim \phi$. Thus we plot $\mathrm{stress}/\phi$ (Fig. \ref{fig:stress-strain}) to reveal effects due to interactions. For passive particles we find that the stress increases faster than $\phi $, whereas for high activity the stress is sub-linear in $\phi$. To understand this observation, note that the $y$ momentum dissipated into the substrate is proportional to the integral of the flow velocity, $\int \langle v_y(x) \rangle dx$. Since the wall is the dominant source of the average momentum, this implies that the stress at the wall is also proportional to $\int \langle v_y(x) \rangle dx$. Thus, if $v_y(x)$ penetrates farther into the bulk, stress at the wall is increased. On the other hand, recall from section \ref{sec-role-of-interactions} that the penetration depth of $v_y(x)$ increases with $\phi$ for passive particles, and decreases with $\phi$ for active particles. The combination of these two effects implies that the stress at the wall should increase faster than $\phi$ for passive particles, and the opposite for sufficiently active particles, consistent with our observations.

%However, above the onset of motility-induced phase separation ($\phi \gtrsim 0.5$ in our simulations), the wall stress decreases with activity. \mfh{Is the last sentence correct? Your sentence was somewhat vague.}

%interactions are enhancing momentum transport away from the boundary, consistent with the discussion in section \ref{sec-role-of-interactions}.
%Similarly, for large enough activity the stress increases more slowly than $\phi $, indicating that indicates that
%interactions are blocking momentum transport away from the boundary. Past $\phi \approx 0.5$ the trend reverses, which coincides with the onset of phase separation.

%For passive particles the stress increases faster than $\phi $, indicating that
%interactions are enhancing momentum transport away from the boundary, consistent with the discussion in section \ref{sec-role-of-interactions}.
%Similarly, for large enough activity the stress increases more slowly than $\phi $, indicating that indicates that
%interactions are blocking momentum transport away from the boundary. Past $\phi \approx 0.5$ the trend reverses, which coincides with the onset of phase separation.

\section{Discussion}

Using Brownian dynamics simulations and kinetic arguments, we have investigated the phenomenology of a boundary-driven active gas in a sheared channel geometry. We find that the nontrivial parts of this phenomenology are confined to a boundary layer characterized by the microscopic length scales $\ell$ (the active persistence length), $\sqrt{D/\xi}$ (the thermal persistence length), and $\sigma$ (the particle diameter). We do not observe spontaneous flow or density inhomogeneities in the bulk for parameters below the onset of motility-induced phase separation.

Within the boundary layer, the mechanisms of momentum transport are dictated by the complex interplay among interparticle interactions, active forces, and boundary driving. Depending on the system parameters, the presence of interactions can either aid or hinder momentum transport. More dramatically,  flow reversal can manifest in the large friction limit due to collision-induced orientational order within the boundary layer.

Although we have rationalized these findings in terms of a simple kinetic picture, it is an open question whether a more systematic theoretical description in terms of appropriate hydrodynamic variables and constitutive relations exists. Our results suggest difficulties in developing this type of description, however. The nontrivial phenomenology is confined to a boundary layer which cannot be mapped onto a generic bulk description in terms of a finite set of hydrodynamic variables and associated constitutive relations.

Besides addressing such general questions, an interesting topic for future work will be to study the effect of phase separation on the shear response. Boundary driving will non-trivially influence the glassy dynamics observed in high density, weakly active particle fluids \cite{Berthier2014,Flenner2016,Mandal2019} and could potentially exhibit phenomenology similar to shear jamming \cite{Bi2011, Peters2016} and discontinuous shear thickening \cite{Fall2015,Brown2014} seen in passive athermal suspensions. Further, given the coupling in active systems between orientational order and flow, it would also be interesting to study the effect of torques at the boundary. Finally, the phenomena observed here will  generally depend on the full distribution function at the boundary, and in particular on the exact form of the driving force. However, our simulations show that the results from section \ref{sec-results} are at least qualitatively robust against variations of the wall force. To further test the generality of this conclusion, it would be interesting to perform the types of analyses we describe here on other externally-driven active matter systems.

{\bf Acknowledgments.}
We acknowledge support from the Brandeis Center for Bioinspired Soft Materials, an NSF MRSEC,  DMR-1420382 (CGW, AB, MFH), and NSF DMR-1149266 and BSF-2014279 (CGW and AB). Computational resources were provided by the NSF through XSEDE computing resources (MCB090163) and the Brandeis HPCC which is partially supported by the Brandeis MRSEC.

%\section*{Appendix: Additional simulation details}

\section*{Appendix: Additional simulation details}

\subsection*{Simulation parameters}

As described in the main text, all simulations contain $10^4$ particles. On the other hand, we adjust the channel dimensions according to the following rules:
\begin{align}
\xi_0 < 0.1 \, \, \, &\rightarrow \, \, \, L = 100 \sigma \\
0.1 \leq \xi_0 < 1 \, \, \, &\rightarrow \, \, \, L = 75 \sigma \\
1 \leq \xi_0 < 10 \, \, \, &\rightarrow \, \, \, L = 50 \sigma \\
\xi_0 \geq 10 \, \, \, &\rightarrow \, \, \, L = 25 \sigma
\end{align}

This enables improved statistics near the wall in simulations with large friction, where a large value of $L$ is not required to exclude finite-size effects.

Obtaining sufficient statistics at some parameter sets requires additional simulations. To improve the quality of the fits in Fig. 2, at least 30 simulations are run in parallel at each data point and the results averaged. Moreover, for several data points in Fig. \ref{fig:phase-diagram} with large friction and near the phase boundary, we average results over 15 independent simulations (in these cases the effect of density on the flow velocity profile is small and susceptible to noise). Finally, we obtain the results in Fig. \ref{fig:flow-reversal} from averages over 100 independent simulations.

% to confirm the statistical significance of the flow reversal phenomenon.

%To check that steady state had been reached by this time, time series of the flow velocity profiles -- measured by averaging over sequential intervals of $10 \tau$ -- were compared with the corresponding long-time velocity profile.

\subsection*{Dilute limit: fits of $\langle v_y  \rangle$}

\begin{figure}
  \includegraphics[keepaspectratio=true,width=0.59\linewidth]{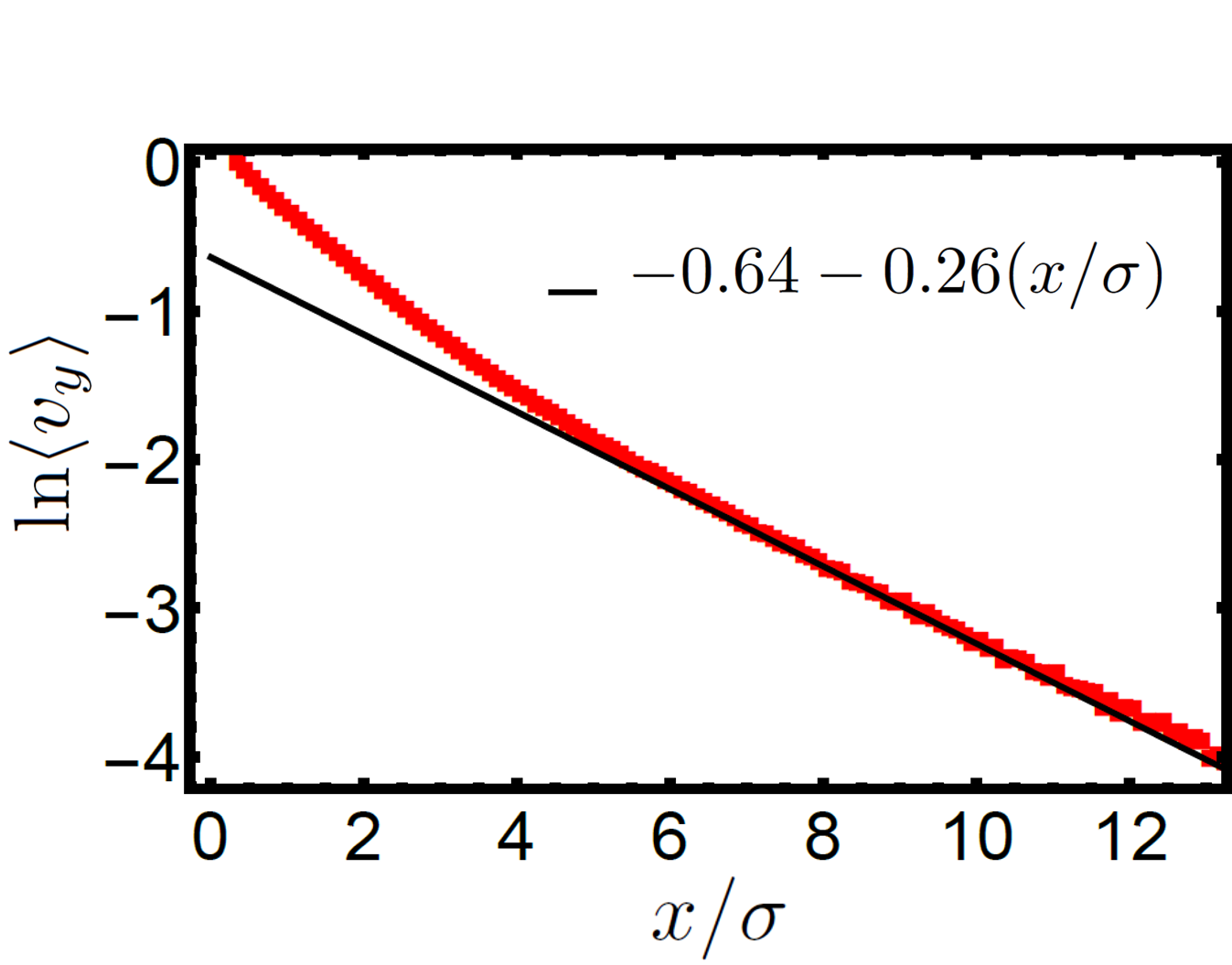}
  \caption{Sample fit of $\langle v_y \rangle$, with $\xi_0 = 1$ and $\ell_0 = 4$.}
  \label{fig:sample_fit}
\end{figure}

For each parameter set we fit $\langle v_y \rangle$ to the form $e^{-x/a}$. The fit is limited to the range $x/\ell_\text{fit}\in(1.5,3.5)$, where  $\ell_\text{fit} = \text{max}(\sqrt{D/\xi}, \ell)$ approximates the distance over which a particle's motion is correlated in a system with no interactions or wall. We select this interval because non-exponential behavior is expected \textit{a priori} for $x \lesssim \ell_\text{fit}$ \cite{Wagner2017}. This prediction is confirmed by our simulation velocity profiles, which exhibit non-exponential behavior in this range. We show an example fit in Fig. \ref{fig:sample_fit} for $\xi_0 = 1$ and $\ell_0 = 4$, which corresponds to $\ell_\text{fit} = 4 \sigma$.

\subsection*{Construction of phase diagram}

We construct the phase diagram in Fig. \ref{fig:phase-diagram} using the following criteria. First, at each point in the $(\xi, \ell)$ space, we measure $\langle v_y \rangle$  for packing fractions $\phi = 0, 0.05, 0.1$, and $0.2$. To quantify how far $\langle v_y \rangle$ penetrates into the bulk, we calculate for each value of $\phi$ the values of $x$ for which $\langle v_y \rangle = 0.3 v_w$ and $\langle v_y \rangle = 0.2 v_w$, denoted as  $x_{0.3}$ and $x_{0.2}$. Finally, we order these quantities with respect to $\phi$. If $x_{0.3}$ and $x_{0.2}$ both increase monotonically with $\phi$, we infer that interactions aid momentum transport, corresponding to green circles in Fig. \ref{fig:phase-diagram}. For the opposite trend, interactions hinder transport, corresponding to red squares. Any other ordering of  $x_{0.3}$ and $x_{0.2}$ is assigned an indeterminate outcome denoted by yellow stars. This might occur due to noise in the measured $\langle v_y \rangle$, or an ambiguous trend in cases where finer variations within a particle diameter of the wall are present.

\bibliographystyle{apsrev4-1}
\bibliography{bib}

\end{document}